\documentclass[aps,prc,preprint,showpacs,showkeys,floatfix]{revtex4}
\usepackage{graphicx}

\newcommand{\bmath}[1]{\mbox{\boldmath${#1}$}}
\newcommand{\ri}[1]{{\bf r}_{#1}}
\newcommand{\brho}{\bmath{\rho}}
\newcommand{\ki}[1]{{\bf k}_{#1}}
\newcommand{\bp}{{\bf p}}
\newcommand{\qi}[1]{{\bf q}_{#1}}

\newcommand{\bkappa}{\bmath{\kappa}}

\newcommand{\bsigma}{\bmath{\sigma}}
\newcommand{\btau}{\bmath{\tau}}
\newcommand{\bpi}{\bmath{\pi}}
\newcommand{\beps}{\bmath{\epsilon}}

\newcommand{\phat}{\bmath{\widehat{\bf p}}}

\newcommand{\rhat}{\bmath{\widehat{\bf r}}}

\newcommand{\btimes}{\bmath{\times}}
\newcommand{\boldT}{\mbox{\boldmath $T$}}

\newcommand{\nablal}{\stackrel{\leftarrow}{\nabla}}
\newcommand{\nablar}{\stackrel{\rightarrow}{\nabla}}

\newcommand{\expup}[1]{e^{#1}}

\newcommand{\EG}{{\textrm{e.g.}}}
\newcommand{\IE}{{\textrm{i.e.}}}
\newcommand{\EA}{{\textit{et al.}}}
\date{\today}
\begin{document}
\vskip1in\vskip1in\vskip1in\vskip1in
\preprint{NT@UW-04-001}
\preprint{FZJ-IKP(TH)-2004-2}
\title{Survey of charge symmetry breaking operators for 
\boldmath$dd\to\alpha\pi^0$}
\author{A. G{\aa}rdestig}
\altaffiliation[Present address:]{Department of Physics and Astronomy, 
Ohio University, Athens, OH 45701}
\email{anders@phy.ohiou.edu}
\affiliation{Department of Physics and Nuclear Theory Center,
Indiana University, Bloomington, IN 47405}

\author{A. Nogga}
\affiliation{Institute for Nuclear Theory, 
University of Washington, Seattle, WA 98195-1550}

\author{C. J. Horowitz}
\affiliation{Department of Physics and Nuclear Theory Center,
Indiana University, Bloomington, IN 47405}

\author{A. C. Fonseca}
\affiliation{Centro Fisica Nuclear, Universidade de Lisboa, 1649-003 Lisboa, 
Portugal}

\author{C. Hanhart}
\affiliation{Institut f\"ur Kernphysik, Forschungszentrum J\"ulich, J\"ulich, 
Germany}

\author{G. A. Miller}
\affiliation{Department of Physics, University of Washington, Seattle, 
WA 98195-1560}

\author{J. A. Niskanen}
\affiliation{Department of Physical Sciences, University of Helsinki, 
Helsinki, Finland}

\author{U. van Kolck}
\affiliation{Department of Physics, University of Arizona, Tucson, 
AZ 85721}
\affiliation{RIKEN BNL Research Center, Brookhaven National Laboratory,
Upton, NY 11973}

\begin{abstract}
The charge-symmetry-breaking amplitudes for the recently observed 
$dd\to\alpha\pi^0$ reaction are investigated.  
Chiral perturbation theory is used  to classify and identify the 
leading-order terms. 
Specific forms of the related one- and two-body tree level diagrams are derived.
As  a first step toward a full calculation,  a few tree-level two-body diagrams 
are evaluated at each considered order, using a simplified set of $d$ and 
$\alpha$ wave functions and a plane-wave approximation for the initial $dd$ state.
The leading-order pion-exchange term is shown to be suppressed in this model
because of poor overlap of the initial and final states.  
The higher-order one-body and short-range (heavy-meson-exchange) amplitudes 
provide better matching between the initial and final states and therefore 
contribute significantly and coherently to the cross section. 
The consequences this might have for a full calculation, with realistic wave 
functions and a more complete set of amplitudes, are discussed.
\end{abstract}


\maketitle

\section{Introduction}
For most practical purposes, hadronic isospin states can be considered as 
charge symmetric, \IE, invariant under a rotation by $180^\circ$ around the 
2-axis in isospin space.
Charge symmetry CS is thus a subset of the general isospin symmetry, charge 
independence CI, which requires invariance under \emph{any} rotation in isospin 
space.
In quantum chromodynamics QCD, CS means that the dynamics are unchanged 
under the exchange of the up and down quarks~\cite{MNS}. 
In the language of hadrons, this symmetry translates into, \EG, the invariance 
of the strong interaction under the exchange of protons and neutrons. 
However, since the up and down quarks do have different masses 
($m_u\neq m_d$)~\cite{wei77,mumd}, the QCD Lagrangian is not charge symmetric 
and neither is the strong interaction of hadrons. 
This symmetry violation is called charge symmetry breaking CSB. 
There is also a contribution to CSB because of the different electromagnetic 
interactions of the up and down quarks. 

Observing the effects of CSB interactions therefore provides a probe of $m_u$ and 
$m_d$, which are fundamental, but poorly known, parameters of the standard model. 
The quantity $m_d$ is larger than $m_u$, causing a specific pattern of mass 
splitting between members of an isospin multiplet~\cite{MNS}. 
In particular, the light quark mass difference causes the neutron to be heavier
than the proton.
If this were not the case, our universe would be very different, as a consequence
of the dependence of Big-Bang nucleosynthesis on the relative abundances of 
protons and neutrons.

Experimental evidence for CSB has been demonstrated in $\rho^0$-$\omega$ 
mixing~\cite{Barkov:ac}, the nucleon mass splitting, the binding-energy 
difference of mirror nuclei such as $^3{\rm H}$ and $^3{\rm He}$~\cite{Nolen:ms}, 
the different scattering lengths of elastic $nn$ and $pp$ 
scattering~\cite{Gabioud}, and in the minute but well-measured difference 
between the proton and neutron analyzing powers of elastic $np$ 
scattering~\cite{Abegg:1986ei}.
A recent theoretical analysis of $\pi N$ scattering data found a small CSB 
effect~\cite{ulfIV}.

Studying the $dd\to\alpha\pi^0$ reaction presents exciting new opportunities 
for developing the understanding of CSB. 
This reaction obviously violates isospin conservation; but more specifically, it
violates charge symmetry since the deuterons and the $\alpha$-particle 
are self-conjugate under the charge-symmetry operator, with a positive 
eigenvalue, while the neutral pion wave function changes sign. 
This reaction could not occur if charge symmetry were conserved, and the 
cross section is proportional to the square of the CSB amplitude.
This is  unique  because all other observations of CSB involve interferences 
with  charge symmetric amplitudes. 
Thus a very clean signal for CSB is obtained through the observation of a 
non-zero cross section.
Furthermore this process has a close connection with QCD because chiral
symmetry plays a dominant role in determining pion-production cross sections.

Lapidus, in 1956~\cite{Lapidus}, was the first to realize that 
the $dd\to\alpha\pi^0$ reaction would  be a useful probe of CSB.
Various experimental groups tried to observe it, but without success~\cite{oldexp}.
After other attempts yielding only upper limits~\cite{Banaigs}, a group at the 
Saturne accelerator in Saclay reported a non-vanishing $dd\to\alpha\pi^0$ 
cross section at $T_d=1.1$~GeV~\cite{Goldzahl}. 
This finding was refuted by members of the same collaboration who argued 
that the putative signal for $\pi^0$ production actually was caused by the 
$dd\to\alpha\gamma\gamma$ background~\cite{FP}. 
The importance of this background was confirmed by calculations of the double 
radiative capture~\cite{DFGW}, using a model based on a very successful treatment 
of the $dd\to\alpha\pi\pi$ reaction at similar energies~\cite{ABC}.
Thus the Saclay $dd\to\alpha\pi^0$ cross section is almost certainly a 
misinterpretation of a heavily-cut smooth $dd\to\alpha\gamma\gamma$ 
background~\cite{DFGW}.

There have been two exciting recent observations of CSB in experiments involving
the production of neutral pions.
Many years of effort have led to the observation of CSB in $np\to d\pi^0$ at 
TRIUMF. 
After a  careful treatment of  systematic errors, the CSB forward-backward
asymmetry of the differential cross section was found to be 
$A_{\rm fb}=[17.2\pm8({\rm stat})\pm5.5({\rm sys})]\times 10^{-4}$~\cite{Allena}.  
In addition, the final experiment at the IUCF Cooler ring has reported 
a very convincing $dd\to\alpha\pi^0$ signal near threshold 
($\sigma=12.7\pm2.2$~pb at $T_d=228.5$~MeV and $15.1\pm3.1$~pb at 231.8~MeV), 
superimposed on a smooth $dd\to\alpha\gamma\gamma$ background~\cite{IUCFCSB}. 
This background is roughly a factor two larger than calculations based on 
Ref.~\cite{DFGW}, but has the expected shape.
The data are consistent with the pion being produced in an $s$-wave,
as expected from the proximity of the threshold ($T_d=225.6$~MeV).

Clearly, these new high-quality CSB experiments demand a theoretical 
interpretation using  fundamental CSB mechanisms.
At momenta comparable to the pion mass, $Q\sim m_\pi$, QCD and its symmetries 
(and in particular CSB) can be described by a hadronic effective field theory 
EFT, chiral perturbation theory $\chi$PT~\cite{ulfreview,birareview}.
This EFT has been extended to pion production~\cite{cpt0,cpt3,Rocha,cpt1,cpt2}
where typical momenta are $Q\sim \sqrt{m_\pi M}$, with $M$ the nucleon mass.
(See also Ref.~\cite{pionprod} where pion production was studied neglecting 
this large momentum in power counting.)
This formalism provides specific CSB effects in addition to the 
nucleon mass difference. 
In particular, there are two pion-nucleon seagull interactions related by chiral 
symmetry to the quark-mass and electromagnetic contributions to the 
nucleon mass difference~\cite{vkiv,wiv}. 

It was demonstrated for the CI reactions $\pi\pi\to\pi\pi$~\cite{Ecker},
$\pi N\to\pi N$~\cite{ulfreview}, and $NN\to NN$~\cite{Epelbaum} 
that the values of the low energy constants can be understood as the low 
energy limit of the exchange of a heavy state. 
This procedure is called the resonance saturation hypothesis.
Within this scheme the other CSB interactions, also caused by the light quark 
mass difference~\cite{vkiv,vKFG}, can be viewed as the low-momentum limit of 
standard meson-exchange mechanisms, such as 
$\pi$-$\eta$-$\eta'$ and $\rho$-$\omega$ mixing.
Determining the various interaction strengths may provide 
significant information about the quark mass difference.
Since these terms contribute to CSB in the reactions 
$np\to d\pi^0$ and $dd\to\alpha\pi^0$ with different weights, 
it is important to analyze both processes using the same framework.

Early calculations of CSB in $np\to d\pi^0$~\cite{chm79,nis99} incorporated most of
the relevant mechanisms, giving an asymmetry --- dominated by $\pi$-$\eta$ mixing 
--- of the order of $-2\times10^{-3}$ for energies near threshold~\cite{nis99}. 
The combined pion-nucleon seagull interactions required by chiral
symmetry generate a larger contribution with the opposite sign~\cite{vKNM},
and provide a prediction for $A_{\rm fb}(np\to d\pi^0)$ (based on a
crude estimate of the strength of the CSB rescattering contribution) that 
was confirmed by the recent experimental observation.
However, the experimental value is in the lower band
of the predicted range of values of $A_{\rm fb}$.

Our aim here is  to provide the  first study of CSB in the near threshold 
$dd\to\alpha\pi^0$ reaction using chiral EFT techniques. 
The effect of $\pi$-$\eta$-$\eta'$ mixing on this reaction
was studied several years ago at $T_d=1.95$ GeV~\cite{cp86}.
Pion production was assumed to be dominated by the production of $\eta$ and 
$\eta'$, followed by $\pi$-$\eta$ or $\pi$-$\eta'$ mixing.
Using phenomenological information on these parameters and on the $\eta$-$\eta'$ 
angle, the cross section was expressed in terms of existing  data  
for the $\eta$ production cross sections. 
This method cannot be used for energies lower than that required to produce 
an $\eta$ meson, and other CSB contributions cannot be evaluated this way. 

It is necessary to  explicitly account for the detailed dynamics of the 
few-nucleon pion-production amplitudes.
Therefore we  will discuss the CSB amplitudes in  the first few orders,
defined according to a chiral counting scheme that provides a general guide to 
the expected importance of different interaction terms. 
Such schemes do not explicitly account for spin-isospin factors, 
for the sometimes poor overlap of wave functions, 
or for the spin and isospin dependence of the wave functions.
We shall see that selection rules resulting from the use of specific wave 
functions and the threshold  kinematics have a strong impact on the
relative importance of particular diagrams.

The fast incoming deuterons ($p\sim460$~MeV/$c$ in the center-of-momentum frame 
c.m.) need to be slowed down to produce an $\alpha$-particle and an $s$-wave 
pion at threshold.
The resulting  large momentum transfer can be transmitted through the 
initial- and final-state interactions or wave function distortions, and through 
the exchange of a particle in the pion-production sub-amplitude.
Only the latter two possibilities will be considered here. 
The complexities of the $dd$ initial state interaction 
will be included in a future study.
Thus, we expect that a pion-production sub-amplitude should preferentially provide 
for momentum sharing between the deuterons, in order to avoid forcing the nucleons 
out into the small, high-momentum tail of the $\alpha$-particle wave function.

Spin, isospin, and symmetry requirements restrict the partial waves
allowed for the $dd\to\alpha\pi^0$ reaction.
In the spectroscopic notation $^{2S+1}\!L_Jl$, where $S$, $L$, $J$ are the 
spin, orbital, and total angular momenta of the $dd$ state and $l$ is the pion
angular momentum, the lowest partial waves are $^3\!P_0s$ and $^5\!D_1p$. 
Hence, production of an $s$-wave pion requires that the initial deuterons be
in a relative $P$-wave, with spins coupled to a spin-1 state, 
coupled together to zero total angular momentum. 
The deuteron spins then need to be flipped, while absorbing the $P$-wave, to 
form the spin-0 state of the helium nucleus. 
The invariant amplitude therefore takes the form $\bp\cdot(\beps_1\btimes\beps_2)$
where $\bp$ is the deuteron relative momentum and $\beps_{1,2}$ are the 
polarization vectors of the initial deuterons.
On the other hand, a $p$-wave pion is produced only when the deuterons are in a
relative $D$-wave, with spins maximally aligned to spin 2, requiring either a 
coupling with $\Delta L=\Delta S=2$ or $D$-states of $d$ or $\alpha$.
This invariant amplitude is of the form 
$\bp\cdot\beps_1\bp\cdot(\beps_2\times\bp_\pi)+
\bp\cdot\beps_2\bp\cdot(\beps_1\times\bp_\pi)$, 
where $\bp_\pi$ is the pion momentum.
Interferences between $s$ and $p$-waves will disappear for any unpolarized 
observable.

In addition to these momentum-sharing and overall symmetry 
considerations, the spin-isospin symmetries of the nucleons in the 
$dd\!:\!\alpha$ system will turn out to be crucial in determining which 
sub-amplitudes can contribute and what possible meson exchanges 
can take place. 
This will be discussed in considerable detail below.

In this first stage we explore the $dd\to\alpha\pi^0$ production process 
using chiral EFT with the simplest deuteron and $\alpha$-particle wave functions,  
and ignoring the effects of initial-state interactions. 
This will give us an initial test of the amplitudes and provide us with the 
framework necessary to establish the ingredients for a full-fledged model.
We are developing a full model, using realistic wave functions and incorporating 
initial-state interactions, along with $\Delta$ admixtures, and the results will 
be reported in forthcoming papers.

The chiral power counting scheme is developed in Sec.~\ref{Ampls}, 
resulting in a list of possible CSB amplitudes.
Our simplified model is presented in Sec.~\ref{sec:Model}.
The relative importance of the amplitudes in this model is investigated in 
Sec.~\ref{Results}.
The paper then concludes in Sec.~\ref{Disc} with a discussion of the results, 
implications for the interpretation of the IUCF experiment, and future prospects.
Some details of the calculation are included in an Appendix.

%
%
\section{CSB operators}
\label{Ampls}
We use the EFT power-counting scheme to classify the 
CSB pion production operators in this section. 
In addition, the specific forms of the tree-level one- and two-body 
operators are derived.
A few unknown low-energy constants LECs appear in the first few orders. 
Since these cannot be determined by symmetry considerations, we use 
phenomenological transition amplitudes to estimate their size.   
The effects of the derived  operators are evaluated using a 
simplified model in Sec.~\ref{sec:Model}. 
This allows us to check that the leading non-vanishing operators of the chiral 
expansion indeed lead to a CSB cross section of the observed order of magnitude.

\subsection{Effective Interactions}
In QCD, the pseudo-Goldstone bosons of spontaneously broken chiral symmetry,
$SU(2)\times SU(2) \to SU(2)$, can be identified with the pions. 
Chiral symmetry then strongly constrains the interactions allowed for
pions with matter, and it is possible to construct a well-defined, convergent
effective field theory for near-threshold pion reactions, namely chiral
perturbation theory. 
Reviews with special emphasis on nucleon systems
are provided in, \EG, Refs.~\cite{ulfreview,birareview}.
The chiral expansion can be adapted to the larger momentum scale inherent in 
pion production in nucleon-nucleon and nucleus-nucleus 
collisions~\cite{cpt0,cpt3,Rocha,cpt1,cpt2}.
The necessary power series may converge (albeit slowly)
for this class of reactions~\cite{cpt1,cpt2}. 
Studies of the $pp\to pp\pi^0$ reaction have shown that the resonance-saturation 
hypothesis does not necessarily lead to couplings of natural size, at least 
for interactions that contribute to the production of $s$-wave pions~\cite{cpt3}.
This issue should be further investigated.

We intend to reproduce the $S$-matrix elements of QCD at momenta much smaller
than the chiral-symmetry-breaking scale,
here identified for simplicity with the nucleon mass $M$.
To do this, the low-energy EFT must contain
all the interactions among pions $\bpi$, nucleons $N$, 
and Delta-isobars $\Delta$, that are allowed by the symmetries of QCD.
For the following, the relevant CI interactions are 
\begin{eqnarray}
\mathcal{L}_{\rm CI} &=& 
         -\frac{1}{4 f_\pi^{2}} 
           N^{\dagger}[\btau \cdot(\bpi\times\dot{\bpi})]N 
        + \frac{g_A}{2f_\pi}
        \left\{N^\dagger\btau\cdot\vec\sigma\cdot N(\vec\nabla\bpi)
        -\frac{1}{2 M}\left[iN^{\dagger}\btau\cdot\dot{\bpi}
        \vec{\sigma}\cdot\vec{\nabla}N + h.c.\right] \right\}
\nonumber\\
        && + \frac{h_A}{2f_\pi}
        \left\{N^\dagger \boldT\cdot
          \vec{S}\cdot\Delta (\vec{\nabla}\bpi) +h.c. 
        -\frac{1}{M}\left[iN^{\dagger}\boldT\cdot\dot{\bpi}
        \vec{S}\cdot\vec{\nabla}\Delta +h.c.\right] \right\}.
\label{CIlag}
\end{eqnarray}
Here the first interaction is the Weinberg-Tomozawa term whose
strength is fixed by chiral symmetry in terms of the
pion decay constant $f_\pi = 92.4$ MeV.
The other terms represent the standard axial-vector couplings
 --- including recoil --- of the pion to the nucleon (with $g_A= 1.26$)
and to the Delta-isobar (with $h_A=2.8$).
Note that $\vec{\sigma}$ and $\btau$ are the usual Pauli matrices 
in spin and isospin space, and $\vec{S}$ and $\boldT$ are the standard
$N\Delta$ spin and isospin transition  matrices, normalized such that
$
  S_{i}S^{+}_{j} =  \frac{1}{3} (2\delta_{ij} - 
              i\varepsilon_{ijk} \sigma_{k}),    
  T_{a}T^{+}_{b}  =  \frac{1}{3} (2\delta_{ab} 
              - i\varepsilon_{abc} \tau_{c}).   
$

Charge symmetry breaking can occur either via exchange of 
a long-wavelength (soft) virtual photon or via short-range interactions.
The former is generated by writing all allowed gauge-invariant interactions
of the photon field.
The latter are represented by local interactions that come either from the quark 
mass difference $m_u-m_d \equiv \epsilon (m_u+m_d)$, or from the exchange of 
short-wavelength (hard) photons (``indirect'' electromagnetic effects), or both.
The relevant CSB interactions are 
\begin{eqnarray}
\mathcal{L}_{\rm CSB} & = & 
        \frac{\delta M}{2}
        N^\dagger\left(\tau_3-\frac{\pi_3\btau\cdot\bpi}
        {2f_\pi^2}\right)N
        +\frac{\bar\delta M}{2}
        N^\dagger\left(\tau_3+
        \frac{\pi_3\btau\cdot\bpi-\bpi^2\tau_3}{2f_\pi^2}\right)N
        \nonumber \\ 
   & &  -\frac{3\delta M}{8M^2}
        \left[N^\dagger\left(\tau_3-\frac{\pi_3\btau\cdot\bpi}
        {2f_\pi^2}\right)\nabla^2 N 
        + (\nabla^2 N)^\dagger\left(\tau_3-\frac{\pi_3\btau\cdot\bpi}
        {2f_\pi^2}\right) N \right]
        \nonumber \\ 
   & &  -\frac{3\bar\delta M}{8M^2}
        \left[N^\dagger\left(\tau_3+
        \frac{\pi_3\btau\cdot\bpi-\bpi^2\tau_3}{2f_\pi^2}\right)\nabla^2 N 
        + (\nabla^2 N)^\dagger\left(\tau_3+
        \frac{\pi_3\btau\cdot\bpi-\bpi^2\tau_3}{2f_\pi^2}\right) N 
        \right]
        \nonumber \\ 
    & & +\frac{1}{4M^2 f_\pi^2}
        N^\dagger
        \left[-\delta M \nabla^2\left(\pi_3\btau\cdot\bpi \right)
        +\bar\delta M \nabla^2\left(\pi_3\btau\cdot\bpi-\bpi^2\tau_3\right)
        \right]N
        \nonumber \\ 
    & & +\frac{1}{2M^2} i \varepsilon_{ijk}
        \left[-\delta M (\partial_i N)^\dagger
        \left(\pi_3\btau\cdot\bpi \right) \sigma_k \partial_j N
        +\bar\delta M (\partial_i N)^\dagger
        \left(\pi_3\btau\cdot\bpi-\bpi^2\tau_3\right)\sigma_k \partial_j N
        \right]
        \nonumber \\ 
    & &
        -\frac{(\beta_1+\bar\beta_3)}{2f_\pi}
        \left\{N^\dagger\vec\sigma N \cdot \vec\nabla \pi_3
        -\frac{1}{2 M}\left[iN^{\dagger}\dot{\pi}_3
        \vec{\sigma}\cdot\vec{\nabla}N + h.c.\right] \right\}        
        +\ldots\ ,
\label{eq:CSBlag}
\end{eqnarray}
where $\delta M = O(\epsilon m_\pi^2/M)$ and $\bar\delta M = O(\alpha M/\pi)$
are, respectively, the quark-mass-difference and electromagnetic contributions 
to the nucleon mass difference, and $\beta_1 = O(\epsilon m_\pi^2/M^2)$ and
$\bar\beta_3 = O(\alpha/\pi)$ are, respectively, the quark-mass-difference and
electromagnetic contributions to the isospin-violating pion-nucleon coupling.
This Lagrangian is consistent with the one from Ref.~\cite{cpt0}, 
with the $\bar\delta M$ term added from Ref.~\cite{vKNM} and the 
pion-nucleon $(\beta_1+\bar\beta_3)$ term from Ref.~\cite{vKFG}.
This Lagrangian is also consistent with that of Ref.~\cite{vKFG}. 
An apparent difference of an overall minus sign arises because Ref.~\cite{vKFG} 
used different signs for the pion field and for $\delta M+\bar{\delta}M$.
The CSB seagull term is consistent with the one used in Ref.~\cite{vKNM}.
These and other CSB EFT interactions were considered in Refs.~\cite{cpt2,vkiv}.

As usual, we have used~\cite{vkiv} naive dimensional analysis to estimate
the strengths of the terms in the Lagrangian, \IE, we have assumed that 
the LECs are of natural size.
In principle, these parameters should be determined using  experimental data. 
We now discuss some of the information we have about them.

The first two terms of Eq.~(\ref{eq:CSBlag}) are the pion-nucleon seagull 
interactions required by chiral symmetry~\cite{vkiv,wiv} and can be described 
as the CSB components of the pion-nucleon $\sigma$-term. 
The strengths are determined by the coefficients $\delta M$ and $\bar\delta M$, 
with their sum related to the nucleon mass splitting: to this order,
\begin{equation}
\delta M+\bar{\delta}M=\Delta M=M_n-M_p=1.29~{\rm MeV}.
\label{eq:M}
\end{equation}
The coefficients are not well-known separately. 
With some assumptions about higher-energy physics, the Cottingham sum rule
can be used to give $\bar\delta M=-(0.76\pm 0.30)$ MeV~\cite{gl}.
It is desirable to determine these parameters without these assumptions.
The $\delta M$, $\bar\delta M$ contribution to other observables 
generally depends on a different combination than that in Eq.~(\ref{eq:M}).
It is difficult to isolate the parameters in $\pi N$ scattering,
so it was suggested~\cite{vKNM} that CSB in pion production
could be used instead.
The forward-backward asymmetry in $np\to d\pi^0$ was shown
to be sensitive to $\delta M-\bar{\delta}M/2$,
but it also depends significantly on $\beta_1+\bar\beta_3$.

The other LECs are not well-known either. 
The pion-nucleon CSB parameter $\beta_1+\bar\beta_3$ is constrained by the 
Nijmegen phase-shift analysis of the $NN$ scattering data~\cite{nijm} to be
$\beta_1+\bar\beta_3= (0\pm 9) \times 10^{-3}$~\cite{vKFG}.
Below we estimate the impact of this interaction following the standard practice 
of neglecting $\bar\beta_3$ and modeling $\beta_1$ by $\pi$-$\eta$ 
mixing~\cite{vKFG}, which is consistent with the bound from $NN$ scattering.

Among the ``$\ldots$'' in Eq.~(\ref{eq:CSBlag}) there are several CSB short-range 
pion--two-nucleon interactions that contribute in the order we will be considering.
One example is
\begin{equation}
        -\frac{(\gamma_1+\bar\gamma_3)}{2f_\pi} \; N^\dagger N \;
        \left\{N^\dagger\vec\sigma N \cdot \vec\nabla \pi_3
        -\frac{1}{2 M}\left[iN^{\dagger}\dot{\pi}_3
        \vec{\sigma}\cdot\vec{\nabla}N + h.c.\right] \right\},
\label{srCSB}
\end{equation}
where we expect that
$\gamma_1 = O[\epsilon m_\pi^2/(f_\pi^2 M^3)]$  and
$\bar\gamma_3 = O[\alpha/(\pi f_\pi^2 M)]$, for the quark-mass-difference and
electromagnetic contributions respectively.
We know very little about the LECs appearing in these short-range 
pion--two-nucleon interactions, and therefore will  model these LECs with various
heavy-meson-exchange HME mechanisms as detailed  below.

\subsection{Power Counting}
It is necessary to order the various amplitudes  
according to the size of their expected contributions to pion production.
There are several strong-interaction scales in the problem, namely,
\begin{itemize}
\item $\chi =p/M\sim \sqrt{m_\pi/M}$, 
the initial c.m. momentum of the deuteron divided by the 
chiral-symmetry-breaking scale (here identified with the nucleon mass $M$),
which we will use as the expansion parameter;
\item $m_\pi/M\sim\chi^2$, where $m_\pi$ denotes the pion mass;
\item $(M_\Delta-M)/M\sim\chi$, with $M_\Delta$ the Delta mass
\footnote{This assignment has no deeper reason than 
the numerical proximity of $\sqrt{m_\pi M}=2.6m_\pi$ and 
$M_\Delta-M=2.1m_\pi$.}
--- the order assignment given is in line with Ref.~\cite{cpt2}; and
\item $\gamma/M\sim\chi^2$, where $\gamma$
is the typical nucleon momentum inside the 
deuteron and the $\alpha$ particle (for simplicity
we will not distinguish between the two).
\end{itemize}
Moreover, the strengths of CSB effects are governed by
\begin{itemize}
\item $\alpha/\pi$, the fine structure constant that appears with every
exchange of a virtual photon, typically with an extra factor
of $\pi$; and
\item  $\epsilon m_\pi^2/M^2$, the factors of $m_u-m_d$
that come from explicit chiral symmetry breaking via quark-mass terms
\footnote{Apart from the pion mass itself, 
the effects of explicit chiral symmetry breaking are typically sub-leading. 
As a consequence, quark-mass isospin breaking is
in general smaller than it would be expected on the basis of 
the quark mass splitting, $\epsilon \sim 1/3$~\cite{vkiv}.}.
\end{itemize}
We discuss the two types of contributions
individually, to first order, in the following subsections.
Second-order effects in $\alpha$ and $\epsilon$ can also be treated,
but are truly small, and ignored  here.

Power counting in systems of two or more nucleons is complicated
by the fact that some diagrams contain small energy denominators,
corresponding to states that differ from initial and/or final states
only by an energy $O(\gamma^2/M)$.
Sub-diagrams that do not contain such enhancements are denoted as irreducible.
Conservation of energy and momentum in pion production
requires that at least one interaction takes place among nucleons
--- before, during, or after the pion emission.
This interaction transfers a momentum of order $p\sim \sqrt{m_\pi M}$.
When such interactions happen before or after pion emission,
they are included in the (high-momentum tail of the) 
initial- or final-state wave function.
In this case we can speak of a ``one-body'' pion-emission operator.
However, in order to compare sub-amplitudes of the same dimensions
and count powers of $\chi$,
we include these interactions as part of the irreducible
pion sub-amplitude. 
The full pion-production amplitude is  ``reducible'', because  it includes 
further initial- and final-state interactions (via the deuteron and 
$\alpha$ wave functions) that transfer momenta of order $\gamma$.

The separation of reducible and irreducible sub-amplitudes
is convenient because it isolates interactions involving
the scale $\chi$ in the irreducible part. 
Power counting for the initial- and final-state interactions 
corresponding to momenta of $O(\gamma)$ can be done in the usual 
way~\cite{birareview}. 
In this first paper, we use simple wave functions
{\it in lieu} of wave functions obtained in EFT.
The needed EFT wave functions may soon be a reality, since chiral three- and 
four-nucleon calculations already exist~\cite{NNN}.

The loop integrals, propagators and vertices
bring factors of momenta, masses, and coupling constants to any given  diagram.
Dimensional analysis  can be used to  express any coupling constant as appropriate 
powers of $M$ times numbers of order 1 (for CI operators) or 
$\epsilon m_\pi^2/M^2$ or $\alpha/\pi$ (for CSB operators).
Some factors, common to all diagrams, are not written explicitly. 
For example, since we study a system of four nucleons that are bound in an 
$\alpha$ particle in the final state, there are always three loops that
are controlled by $\gamma$. 
Thus, all we need to keep explicitly for a $2n$-nucleon operator (in addition
to what can be read from the vertices and propagators directly) is a common 
factor $(p^3/(4\pi)^2)^{(n-1)}$ (here we have only a three-dimensional integral 
because we estimate the measure of a convolution integral with a wave function). 
Therefore explicit factors of  $\gamma$ are not included
explicitly in the  assignments of chiral order.

As stressed in Refs.~\cite{cpt1,cpt2}, the hierarchy of
diagrams is very different for $s$-wave pions and $p$-wave pions.
We here specialize to $s$-wave pion production, 
relevant for the recent IUCF experiment.

\subsubsection{Diagrams proportional to $\epsilon$}
\begin{figure}
\includegraphics{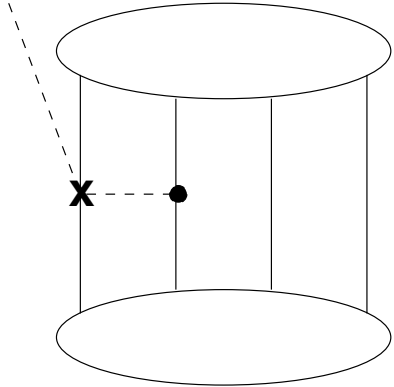}
\caption{Leading order diagram with strong CSB. The cross indicates the
  occurrence of CSB. The dot represents a leading-order CI vertex.}
\label{fig:LOeps}
\end{figure}

At leading order LO there is only one contribution: pion rescattering, where the
CSB occurs through the seagull pion-nucleon terms
linked to the nucleon mass splitting --- see Fig.~\ref{fig:LOeps}, in which 
the leading CI interaction is represented by a dot, and CSB by a cross.
The irreducible part of this diagram is $O[\epsilon m_\pi^2/(f_\pi^3 M p)]$. 
The analogous diagram was identified in Ref.~\cite{vKNM}, using the present 
counting scheme, as giving the dominant contribution to the forward-backward 
asymmetry in $np\to d\pi^0$.
We shall  show that, in the $dd$ induced CSB reaction, selection rules
tend to suppress the rescattering via these seagull terms, 
if initial state interactions are ignored.

\begin{figure*}
\includegraphics{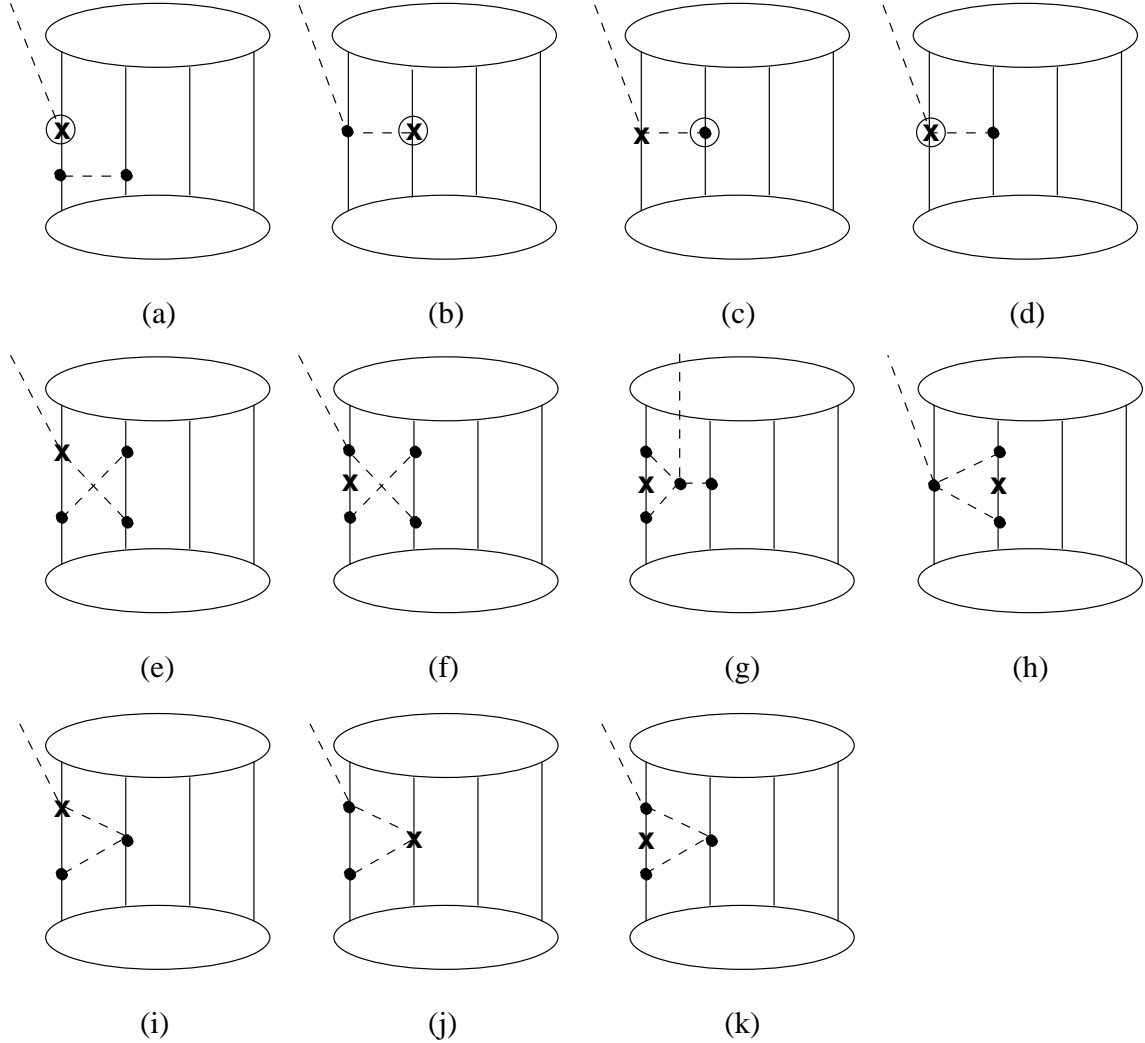}
\caption{NNLO diagrams with strong CSB. 
  Vertices with an additional circle originate from
  sub-leading Lagrange densities. We do not display all possible orderings. }
\label{fig:NNLOeps}
\end{figure*}

There is no next-to-leading order NLO contribution 
(suppressed by just one power of $\chi$). 
At NNLO, however, there are several contributions, displayed in 
Fig.~\ref{fig:NNLOeps}.  
The encircled vertices stem from sub-leading Lagrange densities. 
For example, the sub-leading vertex in diagrams (a) and (b) arises from the 
recoil correction of the CSB $\pi NN$ vertex, the one in diagram (c) denotes 
the recoil correction of the CI $\pi NN$ vertex, and that in diagram (d) 
represents the recoil corrections to the CSB seagulls.
Diagram (b) involves the Weinberg-Tomozawa vertex. 

\begin{figure*}
\includegraphics[scale=0.9]{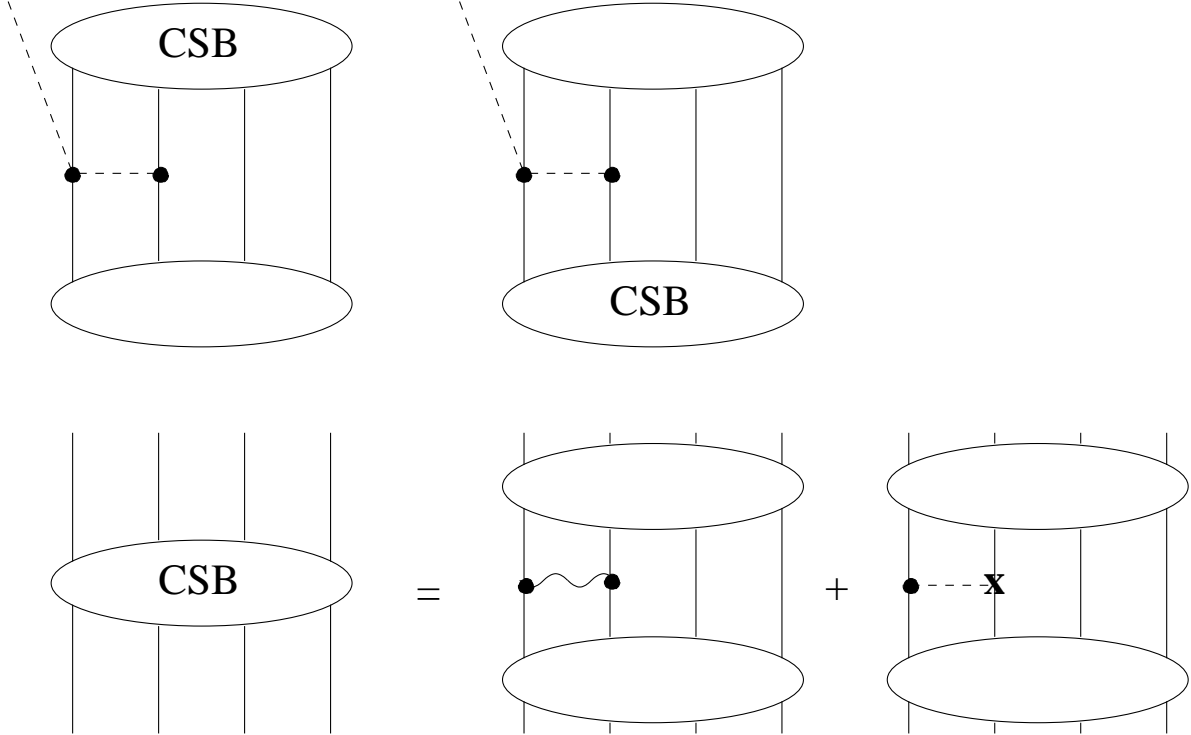}
\caption{The influence of strong and electromagnetic CSB in the initial and
final state.
The wiggly line represents the exchange of a photon, while the
dashed line represents a meson exchange contribution with one CSB vertex.}
\label{fig:wfcsb}
\end{figure*}

Note that diagram (a) can be interpreted as the sandwich of a one-body CSB 
operator between CI initial- and final-state wave functions. 
It is necessary to include the effects of CSB in the wave functions in
addition to the diagrams shown in Fig.~\ref{fig:NNLOeps}.
The easiest way to see this is to compare the size of the LO CSB production 
operator (rescattering via the seagull terms) times the LO CI contribution
to  the $NN$ potential (e.g., one-pion exchange) with  the LO
CI production operator (rescattering via the Weinberg-Tomozawa term)
times the LO CSB contribution to the $NN$ scattering --- assumed
to be one-pion exchange with a CSB coupling on one vertex.
This shows  that CSB in the wave functions should be
significant in a NNLO calculation. 
Typical diagrams are shown in Fig.~\ref{fig:wfcsb}.  

The effects of parity conservation suppress  the influence of CSB in a single
deuteron wave function, but CSB does 
occur in the interactions between the deuterons. 
One such term arises from photon exchange as in Fig.~\ref{fig:wfcsb}.
The dominant CSB contribution in the $\alpha$-particle wave function may be 
expressible in terms of the point radius difference of the neutron and proton
$r_n-r_p$, which can be calculated in microscopic models for few-body systems.
Results of these calculations will be presented in future work.

Loop diagrams appear already at NNLO. 
We display only the topology of these diagrams, 
but it is clearly necessary to include all other orderings. 
A striking feature of the present analysis is that, 
at this order, no counterterms are allowed by the symmetries.
The corresponding counterterms --- the CSB four-nucleon contact
interactions in Eq.~(\ref{eq:CSBlag}), displayed below
in Fig.~\ref{fig:NNXLOeps}(b), appear first at N$^4$LO.
Therefore, those parts of the loops that appear at NNLO
are to be finite. This situation is in complete analogy to 
the CI pion production in nucleon-nucleon collisions discussed in detail in
Ref.~\cite{cpt2}.

Fig.~\ref{fig:NNXLOeps} displays some of the higher-order contributions.
A contribution with an intermediate $\Delta$-isobar, that appears at N$^3$LO,
is shown in diagram (a).
The CSB contact interactions displayed in diagram (b) start to 
contribute at N$^4$LO. 
Their values will be estimated below using phenomenological input.

\begin{figure}
\includegraphics{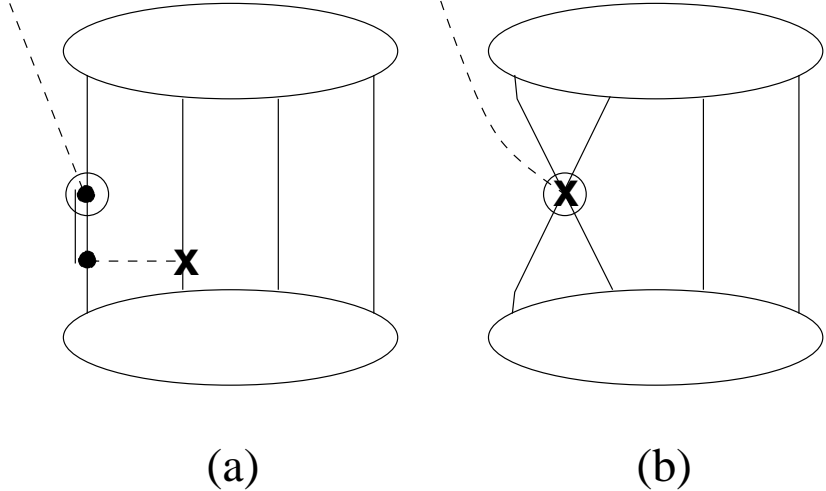}
\caption{Some typical higher-order diagrams with strong CSB. 
  A double line represents a $\Delta$-isobar.
Diagram a) appears at N$^3$LO whereas diagram b) is a N$^4$LO contribution.}
\label{fig:NNXLOeps}
\end{figure}

\subsubsection{Diagrams proportional to $\alpha$}

Electromagnetic contributions can be ordered relative to each
other in exactly the same fashion.
In this case, the LO is $O[\alpha M/(4\pi f_\pi^3 p)]$.
These diagrams contain Coulomb interactions in the initial- or final-state.
In particular, the effects of photon exchange between the initial deuterons, 
followed by production by a strong interaction, could be very important. 
An example of such a term is provided by Fig.~\ref{fig:wfcsb}.

\begin{figure*}
\includegraphics[scale=0.9]{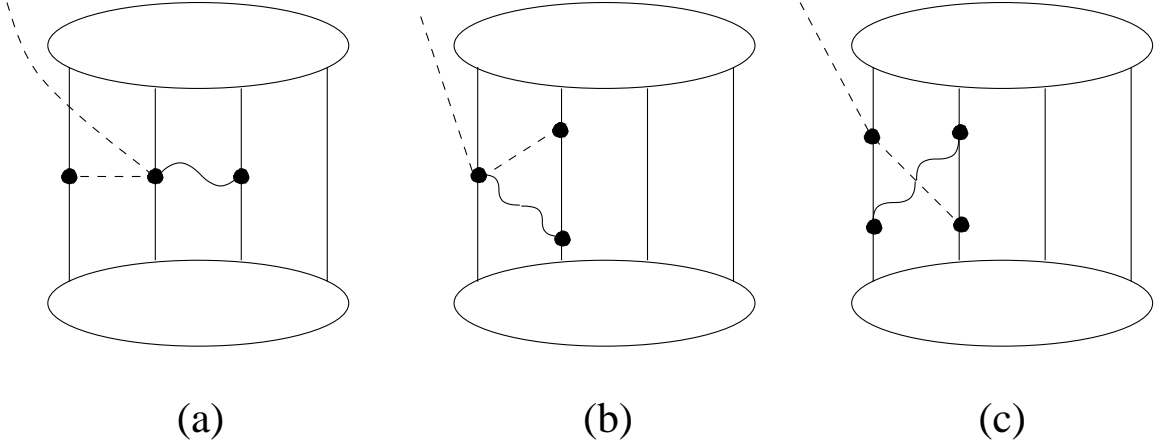}
\caption{NLO diagrams with CSB stemming from soft photons.}
\label{fig:NLOalpha}
\end{figure*}

The NLO electromagnetic diagrams --- suppressed by one power of $\chi$ ---
that contribute to 
CSB in the production operator are shown in Fig.~\ref{fig:NLOalpha}.
It is important to note that in threshold kinematics
(on the two-body level the outgoing nucleons as well as
the produced pion are at rest) the two diagrams (b) and (c)
cancel --- in a realistic calculation we should expect some of
this cancellation effect to survive. 
The three-body diagram (a) should therefore be the one to estimate the 
photon effects in the production operator at this order. 
In addition, higher-order photon couplings in the wave functions 
contribute at this order.

\begin{figure*}
\includegraphics[scale=0.9]{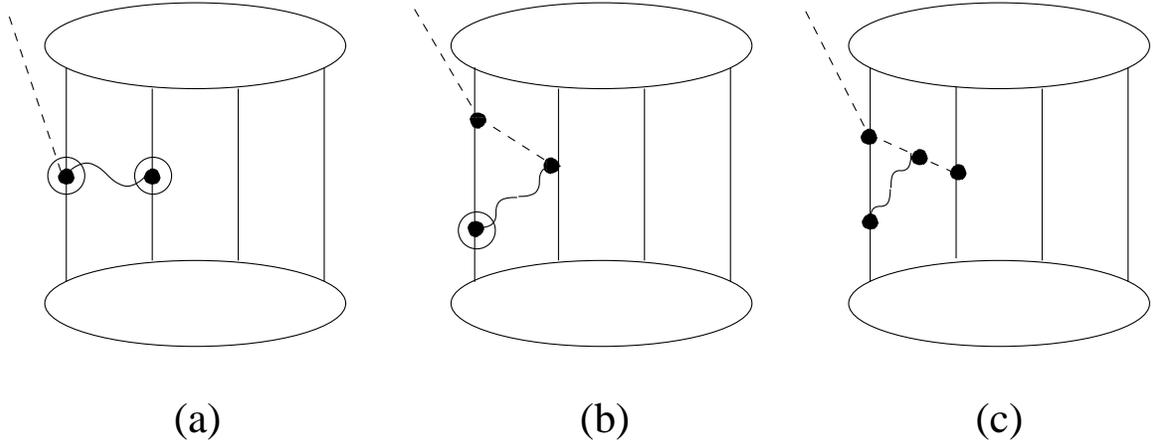}
\caption{NNLO diagrams with CSB stemming from soft photons.}
\label{fig:NNLOalpha}
\end{figure*}

There are various other contributions at NNLO --- see Fig.~\ref{fig:NNLOalpha}.
In what follows we will explicitly calculate the two-body operator that involves 
a photon exchange stemming from gauging the recoil correction to the
$\pi NN$ vertex \footnote{Note that the leading $\gamma\pi NN$ vertex
--- the so-called Kroll-Rudermann term --- couples to the pion charge and
  therefore does not directly contribute to neutral-pion production.},
diagram (a).
This will give us an idea of the relative
importance of soft photons compared to the strong CSB effects.

\subsection{Heavy-Meson  Interactions}
\begin{figure*}
\includegraphics[scale=0.9]{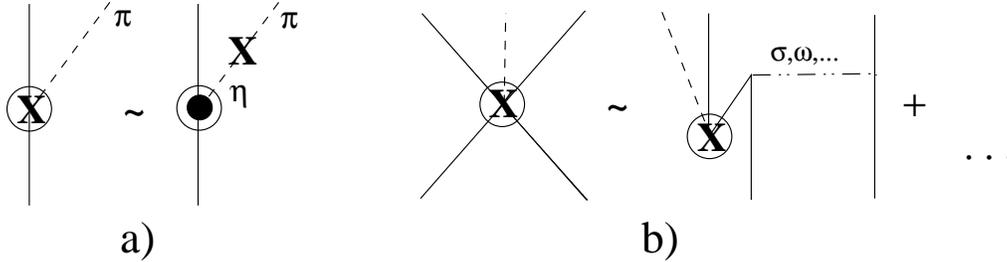}
\caption{Resonance saturation for 
(a) the CSB $\pi NN$ vertex modeled here by $\pi$-$\eta$ mixing and 
(b) the CSB four-nucleon operators.
The ellipsis indicate that additional short-range mechanisms are to be 
included, as discussed in the text.}
\label{ressat}
\end{figure*}

We  assume that   EFT LECs  can be determined using 
the exchange of massive resonances to estimate 
the impact of short-range physics. 
Such an approach was used in CI pion production,
for example, in Refs.~\cite{cpt0,cpt3}.
In principle the counterterms can be determined by  other data, and this 
would eliminate the need for our heavy-meson model.
In the present context, we include the exchanges of the ($\sigma$, $\omega$, 
and $\rho$) mesons depicted in Fig.~\ref{ressat}.

The meson-exchange diagrams can be calculated from the following 
Lagrangian:
\begin{eqnarray}
  \mathcal{L}_{\rm HME} & = & 
  -ig_\eta\bar\psi\gamma_5\psi\eta
  +g_\sigma \bar\psi\psi\sigma
  -g_\omega\bar\psi\gamma_\mu\psi\omega^\mu
  -g_\rho\bar\psi\btau\cdot\left[\gamma_\mu\brho^\mu+
    C_\rho\frac{\sigma_{\mu\nu}}{2M}\partial^\mu\brho^\nu\right]\psi.
\label{eq:LHME}
\end{eqnarray}
Here $\psi$ is the Dirac four-component nucleon field and $\eta$, $\sigma$, 
$\omega^\mu$, $\brho^\mu$ are the meson fields. We use the parameter values in 
Table~\ref{tab:HMEpar} as representative of typical one-boson exchange OBE
models~\cite{EW} and the standard value $C_\rho=6.1$ for the large ratio of
tensor $\sigma_{\mu\nu}\partial^\mu/(2M)$ to vector $\gamma_\mu$
coupling for the $\rho$ meson.
The $\eta$-nucleon coupling $g_\eta$ will be discussed below.

\begin{table}[hbt]
\caption{Table of meson masses and coupling constants.}
\begin{ruledtabular}
\begin{tabular}{ccc}
 & $m$ (MeV/$c^2$) & $\frac{g_{xNN}^2}{4\pi}$ \\
\hline
$\sigma$ & 550 & 7.1  \\
$\omega$ & 783 & 10.6 \\
$\rho$   & 770 & 0.43
\end{tabular}
\end{ruledtabular}
\label{tab:HMEpar}
\end{table}

The photon-nucleon coupling is described by the Lagrangian (up to dimension 5)
\begin{eqnarray}
  \mathcal{L}_\gamma & = & -e\bar\psi\left[\frac{1+\tau^3}{2}\gamma_\mu A^\mu
    +\left(\frac{\lambda_0+\lambda_1\tau^3}{2}\right)
    \frac{\sigma_{\mu\nu}}{2M}\partial^\mu A^\nu\right]\psi,
\end{eqnarray}
where $\lambda_{0,1}=\lambda_p\pm\lambda_n$ and $\lambda_p=1.793$ and 
$\lambda_n=-1.913$ are the proton and neutron anomalous magnetic moments.

\subsection{Explicit form of leading tree-level operators}

We now turn to the explicit form of the leading tree-level two-body
operators, in order to exploit the selection rules. Corresponding expressions 
for the loops as well as the three-body electromagnetic
term mentioned above will be presented in a subsequent publication.

We start with the formally leading mechanism, Fig.~\ref{fig:LOeps},
together with the recoil correction at the pion-nucleon vertex,
Fig.~\ref{fig:NNLOeps}(c).
The pion-exchange operator coming from the seagull terms is
\begin{eqnarray}
        \mathcal{O}_{\pi} & = & 
                \frac{1}{4f_{\pi}^2}
                [\delta M(\btau_i\cdot\btau_j+\tau_i^3\tau_j^3)-
                \bar\delta M(\btau_i\cdot\btau_j-\tau_i^3\tau_j^3)]
        \nonumber \\ && \times \sum_{i\neq j}
        \bsigma_i\cdot\left[(\ki{i}'f^\pi_{ij}-f^\pi_{ij}\ki{i})
            -\frac{q_i^0}{2M}(\ki{i}'f^\pi_{ij}+f^\pi_{ij}\ki{i})\right], \\
\label{eq:Opi}
        f^{\pi}_{ij} & = & \frac{g_A}{2f_{\pi}}
                \frac{{\rm e}^{-\mu r_{ij}}}{4\pi r_{ij}},
\end{eqnarray}
where $\ri{ij}=\ri{i}-\ri{j}$ is the relative coordinate of nucleons $i$ and 
$j$, $\ki{i}=-i\stackrel{\rightarrow}{\nabla}_i$ 
($\ki{i}'=i\stackrel{\leftarrow}{\nabla}_i$) is the initial (final) momentum 
of nucleon $i$, $\qi{i}=\ki{i}'-\ki{i}$ is the momentum transfer
to nucleon $i$ (here symmetrized with the Yukawa factor), and the Yukawa 
parameter $\mu=\sqrt{\frac34}\,m_{\pi}$. 
In our numerical estimates below, we use the value for
$\bar\delta M$ from the Cottingham sum rule, which translates into
$\delta M-\bar\delta M/2=2.4$~MeV~\cite{vKNM}.
In the fixed kinematics approximation for pion production by two nucleons, 
the exchange pion energy $q_i^0=m_\pi/2$~\cite{toy}.

It may be noted that the term from Eq.~(\ref{eq:Opi}), proportional to $\qi{i}$, 
actually gives rise to most of the CSB $s$-wave amplitude in 
$np\rightarrow d\pi^0$~\cite{vKNM}. 
This interferes with CI $p$-wave production. 
On the other hand, the CSB $p$-wave amplitude, arising mainly from the CSB 
one-body operator shown in Eq.~(\ref{eq:one}) or from a CI production operator 
following a CSB initial state interaction, interferes with the CI $s$-wave and 
was about as important in Ref.~\cite{vKNM}, but would be relatively irrelevant 
here in the absence of such an important interference at threshold.

The nucleon recoil term $\sim\frac12(\ki{i}'+\ki{i})$ is smaller, since it is 
suppressed by an additional factor $m_{\pi}/M$.
However, if  the simple deuteron and $\alpha$ wave functions of 
Sec.~\ref{sec:Model} are used, the spin-isospin symmetries prohibit this 
amplitude for nucleons from different deuterons. 
The $\qi{i}$ term will integrate to zero inside a single deuteron, 
leaving the (in-deuteron) recoil as the only allowed contribution. 
Thus the symmetries in this particular model suppress the contribution from 
Fig.~\ref{fig:LOeps}, leaving only Fig.~\ref{fig:NNLOeps}(c):
the seagull amplitude is reduced from LO to NNLO and there is no momentum sharing.
This suppression is expected to be less important once initial state interactions 
are included and realistic wave functions are  used. 

At NNLO there are various other contributions.
The one-body operator, Fig.~\ref{fig:NNLOeps}(a), is 
\begin{eqnarray}
        \mathcal{O}_{1} & = & \frac{\beta_1}{2f_\pi} \sum_{i}\bsigma_i\cdot 
        \left(\qi{i}-\frac{\omega}{2M}(\ki{i}'+\ki{i})\right) \rightarrow 
        \Lambda_1\frac12\sum_i\bsigma_i\cdot(\ki{i}'+\ki{i}), \label{eq:one} \\
        \Lambda_1 & = & -\frac{\beta_1}{2f_\pi}\frac{\omega}{M}.
\end{eqnarray}
The $p$-wave $\qi{i}=-\bp_\pi$ term is suppressed in the threshold regime 
considered. 
In addition, it is not allowed in our plane wave approximation, 
since it lacks the tensor coupling required for the $^5D_1p$ transition. 
The $s$-wave recoil term is allowed, albeit suppressed by a factor $\omega/M$, 
hence the parameter $\Lambda_1$. This $s$-wave term is NNLO. 

The isospin-violating $\beta_1$ is here modeled~\cite{vKFG}
by $\pi$-$\eta$ mixing [see Fig.~\ref{ressat}(a)],
\begin{equation}
\beta_1=\bar{g}_\eta\langle\pi^0|H|\eta\rangle/m_\eta^2, 
\end{equation}
where $\bar{g}_\eta=g_\eta f_\pi/M=0.25$ is the $\eta NN$ coupling constant
and $\langle\pi^0|H|\eta\rangle=-4200$~MeV$^2$ the $\pi$-$\eta$--mixing matrix 
element~\cite{mesmix}. 
The value of $\bar{g}_\eta$ corresponds to $g_\eta^2/4\pi=0.51$, 
similar to the small values found from photo-production 
experiments~\cite{getasmall}. 
However, other values, based on hadronic experiments, are as high as 
$g_\eta^2/4\pi=3.68$~\cite{geta1} or 2--7 for the OBE 
parameterizations of the Bonn potentials~\cite{geta2}. 
The CD-Bonn OBE potential assumes a vanishing value for $g_\eta$, since 
in the full Bonn model no explicit $\eta$ contribution was required by the 
NN data~\cite{geta3}.
Furthermore, the value of the $\pi$-$\eta$--mixing matrix element is uncertain.
With our particular choice we get $\beta_1=-3.5\times10^{-3}$~\cite{vKFG}. 
Using $g_\eta^2/4\pi=3.68$ and $\langle\pi^0|H|\eta\rangle=-5900$~MeV$^2$, as 
done in Ref.~\cite{vKNM}, gives $\beta_1=-1.2\times10^{-2}$.

One important issue is the relative sign of this contribution,
which is apparently not determined experimentally.
The sign given above is consistent with $SU(3)\times SU(3)$ chiral
perturbation theory,
which can be formulated in terms of a pseudoscalar octet 
$\pi_a$ and a baryon octet. 
The sign of the $\pi_3$-$\pi_8$ mixing is, in leading order, fixed by $m_u-m_d$. 
The interactions of $\pi_3$ and $\pi_8$ with the nucleon are determined by 
the standard weak couplings $D$ and $F$, which are fixed in weak decays.
With our definitions of $g_A$, $\bar g_\eta$, and $\beta_1$ given above
and the values of $D$ and $F$ given, e.g., in Ref.~\cite{roxanne},
we find $g_A>0$ if we define $\pi_3 =\pi^0$,
$g_\eta >0$ and $\langle \pi^0 |H|\eta\rangle <0$, so that $\beta_1<0$.
This conclusion holds, as it should,
regardless of the sign definition of $\eta$, 
that is, whether one takes $\eta$ as $\pi_8$ or $-\pi_8$. 

Fig.~\ref{fig:NNLOeps}(b) represents the process where a CSB one-body 
operator produces a charged pion which then changes into a neutral pion 
as it re-scatters on another nucleon via the CI  Weinberg-Tomozawa term.
This contribution is small in $dd\to\alpha\pi^0$, since the isospin 
couplings force the pion exchange to occur inside one of the deuterons. 
This is a situation very similar to the seagull CSB terms, 
which was discussed above, but with a smaller coefficient.
Note that a similar diagram where the exchanged pion is neutral is also 
small, since the on-shell $\pi^0N\to\pi^0N$ amplitude receives contributions 
only at one order higher than that from the Weinberg-Tomozawa term.
Since the operator in Fig.~\ref{fig:NNLOeps}(d) is a relativistic 
correction to the leading order pion rescattering, it has exactly the same 
spin-isospin structure (except its last term) as can be seen in 
Eq.~(\ref{eq:CSBlag}). 
Thus its first few terms are also confined to in-deuteron exchanges and since 
they are already suppressed by two orders $(k_i/M)^2\sim m_\pi/M$, these terms 
are negligible.
The last $\delta M$/$\bar\delta M$ term has an extra Pauli spin matrix 
and can possibly be important since this may allow for momentum sharing. 
However, this term always includes the momentum of a final nucleon, 
which is very small near the pion threshold and this NNLO amplitude 
is likely to be suppressed as well.
We will not consider these operators any further.

The pion loops in Fig.~\ref{fig:NNLOeps}(e-k) represent long-range, 
non-analytic contributions as well as short-range, analytic effects.
The latter cannot be separated from the short-range contributions of 
Fig.~\ref{fig:NNXLOeps}(b), originating from a four-nucleon--pion 
CSB contact interaction. 
In this first study, we limit ourselves to an estimate of these effects 
via resonance saturation from various heavy-meson exchange currents HMECs
--- see Fig.~\ref{ressat}(b).
In the case of the $pp\to pp\pi^0$ reaction, heavy-meson exchanges 
involving the creation of a nucleon--anti-nucleon pair (z-graphs) were 
shown to be important for the total (CI) cross section
near threshold~\cite{leeriska,chucketal,cpt3}.
These exchanges correspond to contact interactions in the
EFT~\cite{cpt0,cpt3}.
Here, we include the analogous CSB interactions where CSB occurs 
in the pion emission or in the meson exchange. 

The HME two-body operators are derived directly from a low-energy reduction 
of the Feynman rules for the HME Lagrangian Eq.~(\ref{eq:LHME}).
This gives the $\sigma$-meson--exchange two-body operator
\begin{eqnarray}
        \mathcal{O}_{\sigma} & = & \Lambda_1
        \frac{1}{2} \sum_{i\neq j}\bsigma_i\cdot
        (\ki{i}'f^\sigma_{ij}+f^\sigma_{ij}\ki{i}), \label{eq:Os} \\
        f^{\sigma}_{ij} & = & \frac{g_\sigma^2}{4\pi M}
                \frac{{\rm e}^{-m_\sigma r_{ij}}}{r_{ij}},
\label{eq:sigma}
\end{eqnarray}
where only the symmetrized recoil term has been used.
Note that the sum is over $i\neq j$ rather than $i<j$. 

The $\omega$-exchange two-body operator is
\begin{eqnarray}
        \mathcal{O}_\omega & = & -\Lambda_1\frac{1}{2}\sum_{i\neq j}
                \left[\bsigma_j\cdot(\ki{i}'f^\omega_{ij}+f^\omega_{ij}\ki{i})
                +i(\bsigma_i\btimes \bsigma_j)\cdot 
                (\ki{j}'f^\omega_{ij}-f^\omega_{ij}\ki{j}) \right], \\
        f^\omega_{ij} & = & \frac{g_\omega^2}{4\pi M}
                \frac{{\rm e}^{-m_\omega r_{ij}}}{r_{ij}}.
\end{eqnarray}
Note this has an overall minus sign and $\bsigma_j$ instead of $\bsigma_i$ 
compared to $\mathcal{O}_\sigma$.  
Finally there is a new term involving the momentum transferred to nucleon $j$. 

The $\rho$-exchange two-body operator is
\begin{eqnarray}
\mathcal{O}_\rho & = & -\Lambda_1
        {1\over 2} \sum_{i\neq j} \btau_i\cdot\btau_j\left[\bsigma_j \cdot 
        (\ki{i}'f^\rho_{ij}+f^\rho_{ij}\ki{i}) + i 
        (1+C_\rho)(\bsigma_i\btimes\bsigma_j)\cdot
        (\ki{j}'f^\rho_{ij}-f^\rho_{ij}\ki{j})\right], \\
        f^\rho_{ij} & = & \frac{g_\rho^2}{4\pi M}
                \frac{{\rm e}^{-m_\rho r_{ij}}}{r_{ij}}.
\end{eqnarray}
The $\rho$ HMEC is of order of the small vector times the large tensor 
coupling constant and has no contributions of order of the tensor coupling 
squared. 

The $\rho$-$\omega$--mixing two-body operator is
\begin{eqnarray}
        \mathcal{O}_{\rho-\omega} & = & -\Lambda_{\rho-\omega}
                {1\over 2} \sum_{i\neq j} \Bigl\{(1+\tau_i^3\tau_j^3)
        \bsigma_j\cdot(\ki{i}'f^{\rho\omega}_{ij}+f^{\rho\omega}_{ij}\ki{i}) \\
  && + i[1+\tau_i^3\tau_j^3(1+C_\rho)](\bsigma_i\btimes\bsigma_j)\cdot
        (\ki{j}'f^{\rho\omega}_{ij}-f^{\rho\omega}_{ij}\ki{j}) \Bigr\}, \\
        \Lambda_{\rho-\omega} & = & -\frac{g_A}{2f_\pi}\frac{\omega}{M}
       \left(\frac{\langle\rho|H|\omega\rangle}{m_\omega^2}\right), \\
        f^{\rho\omega}_{ij} & = &
                \frac{g_\rho g_\omega}{4\pi Mr_{ij}}
                \frac{m_\omega^2}{m_\omega^2-m_\rho^2}
                ({\rm e}^{-m_\rho r_{ij}}-{\rm e}^{-m_\omega r_{ij}}),
\label{eq:Orhomega}
\end{eqnarray}
where the $\rho$-$\omega$ mixing is given by 
$\langle\rho|H|\omega\rangle=-4300$~MeV$^2$~\cite{mesmix}.
A somewhat smaller number ($\langle\rho|H|\omega\rangle=-3500\pm300$~MeV$^2$) was 
obtained in a more recent analysis~\cite{Gardner:1997ie}.
The isospin-independent part of this $\rho$-$\omega$ operator 
is only of the order of the small $\rho$ vector coupling.  
Note, however, that there is a $\tau_i^3\tau_j^3$ term that involves the large 
$\rho$ tensor coupling.  

At momenta much smaller than the heavy-meson masses these HMECs
are equivalent to short-range pion--two-nucleon contact interactions,
with specific values for the LECs.
For example, the $\sigma$ mechanism [Eq.~(\ref{eq:sigma})] goes into the 
interaction shown in Eq.~(\ref{srCSB})
with $\gamma_1$ given by $\beta_1 g_\sigma^2/(4\pi m_\sigma^2 M)$,
which is consistent with the naive dimensional estimates.

In addition, we need to consider contributions from soft photons.  
There is a Coulomb interaction and a magnetic interaction 
(Fig.~\ref{fig:wfcsb}), and a three-body term [Fig.~\ref{fig:NLOalpha}(a)]. 
As a first  estimate, we shall compute the lowest order two-body diagram with a
photon. This appears at NNLO and is shown  in Fig.~\ref{fig:NNLOalpha}(a).

The soft photon exchange gives a structure very similar to that of
$\rho^0$-$\omega$ mixing:
\begin{eqnarray}
  \mathcal{O}_{\gamma} & = & -\Lambda_{\gamma}
      {1\over 2} \sum_{i\neq j} \Bigl\{(1+\tau_i^3\tau_j^3)
      \bsigma_j\cdot(\ki{i}'f^{\gamma}_{ij}+f^{\gamma}_{ij}\ki{i}) \nonumber \\
  && + i[1+\lambda_0+(1+\lambda_1)\tau_i^3\tau_j^3](\bsigma_i\btimes\bsigma_j)
          \cdot(\ki{j}'f^{\gamma}_{ij}-f^{\gamma}_{ij}\ki{j}) \Bigr\}, \\
      \Lambda_{\gamma} & = & \frac{1}{4}\frac{g_A}{2f_\pi}\frac{\omega}{M}, \\
       f^{\gamma}_{ij} & = & \frac{\alpha}{Mr_{ij}}.
\end{eqnarray}
Note that the structure of this term is a consequence of gauge invariance, and 
this is why no new unknown parameters are introduced.

%
%
\section{Simplified model}
\label{sec:Model}
\begin{figure}[t]
\scalebox{0.5}{\includegraphics{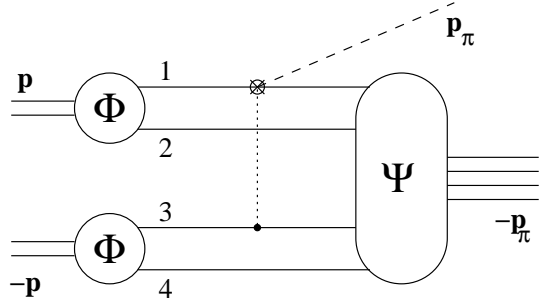}}
\caption{Feynman diagram for pion production in the $dd\to\alpha\pi^0$ 
reaction, indicating the labeling of nucleons and defining basic kinematic
variables.}
\label{fig:ddapi0}
\end{figure}

The interferences and relative importance of the CSB amplitudes of the previous
section can be estimated in a simplified model, using a plane-wave 
approximation and the simplest possible $d$ and $\alpha$ bound-state wave 
functions, those of a Gaussian form. 
A Feynman diagram for this model can be drawn as in Fig.~\ref{fig:ddapi0}.
Assuming spatially-symmetric bound-state wave functions, the
invariant amplitude is given by
\begin{eqnarray}
        \mathcal{M} & = & \int d^3r d^3\!\rho_1 d^3\!\rho_2\,
                        \langle A|\mathcal{O}|DD\rangle, \\
 |A\rangle & = & \sqrt{2E_\alpha}\Psi_\alpha(r,\rho_1,\rho_2)|\alpha\rangle, \\
        |DD\rangle & = & \sqrt{s}\,\Phi_d(\rho_1)\Phi_d(\rho_2)|dd\rangle,
\end{eqnarray}
where $\Psi_{\alpha}$ and $\Phi_d$ are the spatial parts of the 
$\alpha$-particle and deuteron bound-state wave functions, and $s=4E_d^2$ is 
the total c.m.\ energy squared. 
The ket vectors $|\alpha\rangle$ and $|dd\rangle$ contain the fully 
anti-symmetrized spin and isospin wave functions.
Because of the symmetry requirements, the plane-wave $dd$ scattering wave
function is included in $|dd\rangle$ as given by Eqs.~(\ref{eq:dd}) and 
(\ref{eq:ddexp}) below. 
The invariant amplitude can  then be written as
\begin{eqnarray}
        \mathcal{M} & = & \sqrt{2E_{\alpha}s}\int d^3r d^3\rho_1 d^3\rho_2
        \Psi^{\dagger}_{\alpha}(r,\rho_1,\rho_2)
        \langle\alpha|\mathcal{O}|dd\rangle\Phi_d(\rho_1)\Phi_d(\rho_2),
\end{eqnarray}
where $\langle\alpha|\mathcal{O}|dd\rangle$ contains all the spin-isospin 
couplings of the nucleons and the pion production operator $\mathcal{O}$.

The wave functions are expressed in terms of the (2+2) Jacobian coordinates
\begin{eqnarray}
        {\bf R} & = & \frac{1}{4}(\ri{1}+\ri{2}+\ri{3}+\ri{4})\ (\equiv 0\  
                {\rm in\ c.m.}), \nonumber \\
        {\bf r} & = & \frac{1}{2}(\ri{1}+\ri{2}-\ri{3}-\ri{4}), \nonumber \\
        \brho_1 & = & \ri{1}-\ri{2}, \nonumber \\
        \brho_2 & = & \ri{3}-\ri{4},
\end{eqnarray}
with the corresponding momenta
\begin{eqnarray}
  {\bf K} & = & \ki{1}+\ki{2}+\ki{3}+\ki{4}\ (\equiv 0\ 
    {\rm in\ c.m.}), \nonumber \\
  {\bf k} & = & \frac{1}{2}(\ki{1}+\ki{2}-\ki{3}-\ki{4})= 
    \frac{1}{2}({\bf p_1-p_2})\ (\equiv {\bf p}\ {\rm in\ c.m.}), \nonumber \\
  \bkappa_1 & = & \frac{1}{2}(\ki{1}-\ki{2}), \nonumber \\
  \bkappa_2 & = & \frac{1}{2}(\ki{3}-\ki{4}), 
\end{eqnarray}
defined so that 
$\sum_i \ki{i}\cdot{\bf r}_i={\bf K}\cdot{\bf R}+\ki{}\cdot{\bf r}+
\bkappa_1\cdot\brho_1+\bkappa_2\cdot\brho_2$. The Jacobians are equal 
to unity in both representations.

The Gaussian functions that represent the ground state wave functions
are  explicitly expressed in these coordinates 
 using $\sum_{i<j}(\ri{i}-\ri{j})^2=4\ri{}^2+2\brho_1^2+2\brho_2^2$;
\begin{eqnarray}
        \Psi_{\alpha}(r,\rho_1,\rho_2) & = & 
                \frac{8}{\pi^{9/4}\alpha^{9/2}}
                \exp\left[-\frac{1}{\alpha^2}
                \left(2r^2+\rho_1^2+\rho_2^2\right)\right], \\
        \Phi_{d}(\rho) & = & \frac{1}{\pi^{3/4}\beta^{3/2}}
        \exp\left(-\frac{1}{2\beta^2}\rho^2\right),
\end{eqnarray}
where the parameter values $\alpha=2.77$~fm and $\beta=3.189$~fm are derived 
from measured $\alpha$ and $d$ rms point radii;
$\langle r_\alpha^2\rangle^{1/2}=1.47$~fm and 
$\langle r_d^2\rangle^{1/2}=1.953$~fm~\cite{radii}.

Since we have assumed that the orbital parts of the wave functions are 
symmetric under the exchange of any pair of nucleons, we may   define the 
initial- and final-state spin-isospin wave functions as
\begin{eqnarray}
        |\alpha\rangle & = & \frac{1}{\sqrt{2}}\left\{ 
        \left((1,2)_1,(3,4)_1\right)_0\left[[1,2]_0,[3,4]_0\right]_0 -
       \left((1,2)_0,(3,4)_0\right)_0\left[[1,2]_1,[3,4]_1\right]_0 \right\},
        \label{eq:alpha} \\
        |dd\rangle & = & \frac{1}{\sqrt{3}}\left(1-P_{23}-P_{24}\right)
         |d_{12}d_{34}\rangle,
        \label{eq:dd}\\
        |d_{12}d_{34}\rangle & = & ((1,2)_1,(3,4)_1)_S[[1,2]_0,[3,4]_0]_0
                \frac{1}{\sqrt{2}}\left(\expup{i\bp\cdot\ri{}}+
                (-)^L\expup{-i\bp\cdot\ri{}}\right),
        \label{eq:ddexp}
\end{eqnarray}
where $(i,j)_s$ ($[i,j]_T$) are the spin (isospin) Clebsch-Gordan 
couplings, with magnetic quantum numbers suppressed, for nucleons, or nucleon 
pairs, $i$ and $j$ coupling to spin $s$ (isospin $T$). 
Here, $P_{ij}$ is the permutation operator of the indicated nucleons. 
The symmetry requirements for the exchange of the deuterons are represented by 
the (orbital-angular-momentum dependent) combination of plane waves in 
Eq.~(\ref{eq:ddexp}), with ${\bf p}$ as the deuteron relative momentum.
Even though the expression for the $\alpha$ state seems to
single out a (12)+(34) configuration, it is indeed 
fully anti-symmetric in all indices. 
This particular form is used because it closely matches the form of
the initial-state wave functions, simplifying the evaluation of the 
spin-isospin summations in the matrix element. 
The $dd$ wave function can in practice be simplified to
\begin{eqnarray}
        |dd\rangle & = & \sqrt{6}\,((1,2)_1,(3,4)_1)_S[[1,2]_0,[3,4]_0]_0
                (2L+1)i^Lj_L(pr)P_L(\phat\cdot\rhat),
\end{eqnarray}
since each of the three terms in Eq.~(\ref{eq:dd}) gives identical contributions
to the matrix element, and $\expup{i\bp\cdot\ri{}}+(-)^L\expup{-i\bp\cdot\ri{}}$
reduces to $2(2L+1)i^Lj_L(pr)P_L(\phat\cdot\rhat)$ for any particular partial 
wave. 

We may obtain selection rules for the CSB amplitudes that can contribute by 
comparing this expression for the deuterons with the $\alpha$-particle
wave function.  
It is clear that matching the first term of $|\alpha\rangle$ involves no  
nucleon spin or isospin flips, but to match the second term, the spin and isospin
of two nucleons (one from each deuteron) need to be flipped simultaneously. 
Of course, the overall spin has to change in both cases.

In an explicit and straightforward representation, the above spin-isospin 
wave functions can be written as
\begin{eqnarray}
        |\alpha\rangle & = & \frac{1}{2\sqrt{6}}\left\{
\left[\uparrow\uparrow\downarrow\downarrow+\downarrow\downarrow\uparrow\uparrow
  -\frac{1}{2}(
  \uparrow\downarrow\uparrow\downarrow+\downarrow\uparrow\downarrow\uparrow+
  \uparrow\downarrow\downarrow\uparrow+\downarrow\uparrow\uparrow\downarrow)
  \right](pnpn+npnp-pnnp-nppn) \right. \nonumber \\ 
& - & \left.
  (\uparrow\downarrow\uparrow\downarrow+\downarrow\uparrow\downarrow\uparrow
  -\uparrow\downarrow\downarrow\uparrow-\downarrow\uparrow\uparrow\downarrow)
  \left[ppnn+nnpp-\frac12(pnpn+npnp+pnnp+nppn)\right] \right\}, 
\end{eqnarray}
\begin{eqnarray}
        |d_{12}d_{34}10\rangle & = & \frac{\sqrt{6}}{2\sqrt{2}}
(\uparrow\uparrow\downarrow\downarrow-\downarrow\downarrow\uparrow\uparrow)
(pnpn+npnp-pnnp-nppn)\,3ij_1(pr)P_1(\phat\cdot\rhat),
\end{eqnarray}
where the arrows indicate spin projections and $p/n$ proton and neutron 
isospin states.
Note that for the $dd$ state, only the spin-1, $m_S=0$ state is given.
These expressions can then be used together with the Pauli matrices of the 
pion production amplitudes to find the formulas for the matrix elements.

In the normalization used here, the spin-averaged cross section 
(for $s$-wave pions) is given by
\begin{equation}
        \sigma = \frac{1}{16\pi s}\frac{p_{\pi}}{p}
                \frac{1}{9}\sum_{\rm pol.}|\mathcal{M}|^2,
\end{equation}
where the summation is over the deuteron polarizations.

The CSB operators can now be evaluated in this model and studied in more 
detail.
We will start with the simplest operator (\IE, the one-body term) and use 
its matrix element as a reference point for the values  of the other amplitudes. 

\subsection{One-body operator} 
The one-body amplitude is strongly favored by the symmetries of initial and final 
states because all of the nucleons contribute coherently to the cross section. 
However, it does not provide momentum sharing between the deuterons and is 
hence dependent on the shape of the high-momentum tail of the 
$\alpha$-particle wave function. 
The matrix element for this operator is
\begin{eqnarray}
        \mathcal{M}_1 & = & -i\frac{\Lambda_1}{2}
        \bp\cdot(\beps_1\btimes\beps_2)4\mathcal{W}_1, \\
        \mathcal{W}_1 & = & \sqrt{2E_\alpha s}\int d^3rd^3\rho_1d^3\rho_2 
                \Psi_\alpha^\ast j_0(pr)\Phi_1\Phi_2,
\end{eqnarray}
where $\beps_1$ and $\beps_2$ are the polarization vectors of the initial
deuterons and the factor of 4 arises from the sum over all nucleons. 
Thus the spin-momentum structure of the $^3\!P_0s$ partial wave has been 
separated from the dimensionless form factor $\mathcal{W}_1$. 
For Gaussian wave functions this matrix element and the corresponding 
cross section can be calculated analytically. They are
\begin{eqnarray}
        \mathcal{M}_1 & = & -i\frac{\Lambda_1}{2}
             4\bp\cdot(\beps_1\btimes\beps_2)
                \frac{32\pi^{3/4}\sqrt{E_\alpha s}\,\alpha^{9/2}\beta^3}
             {(\alpha^2+2\beta^2)^3}\exp\left(-\frac{\alpha^2p^2}{8}\right), \\
        \sigma_1 & = & \frac{512\sqrt{\pi}}{9}\Lambda_1^2p_\pi p
                \frac{E_\alpha\alpha^9\beta^6}{(\alpha^2+2\beta^2)^6}
                \exp\left(-\frac{\alpha^2p^2}{4}\right),
\label{sig1}\end{eqnarray}
where the exponential stems from the Fourier transform of the $\alpha$-particle
wave function and reflects the dependence on its high-momentum tail.
We will use this one-body estimate as the benchmark for the calculations
of more complicated amplitudes, and also for the full calculation using 
realistic wave functions.

\subsection{Meson-exchange operators}
Although the seagull amplitude is leading order in $\chi$PT, it is suppressed in 
our plane wave treatment of the $dd\to\alpha\pi^0$ reaction because of the 
combination of two $\tau$ matrices and one $\sigma$ matrix, which gives a poor 
match of the initial and final states in our simplified model.
Thus, the pion exchange is allowed only between nucleons from the same 
deuteron, forbidding an advantageous momentum sharing between the deuterons.
In addition, the $\qi{i}$ term vanishes, leaving the recoil term 
$(\omega/M)\ki{i}$ as the only contribution. 
This term is NNLO.
The pion-exchange matrix element is
\begin{eqnarray}
        \mathcal{M}_\pi & = & -i\frac{\Lambda_1}{2}
                \bp\cdot(\beps_1\btimes\beps_2)4\mathcal{W}_\pi, \\
        \mathcal{W}_\pi & = & \mathcal{W}_1\frac{\Lambda_\pi}{\Lambda_1}
                \langle f_{12}^\pi\rangle, \\
        \Lambda_\pi & = & 
        \frac{\left(\delta M-\frac{1}{2}\bar\delta M\right)}{f_\pi^2}
                \frac{\omega}{M}, \\
        \langle f_{12}^\pi\rangle & = & 
                        \frac{1}{\mathcal{W}_1} \sqrt{2E_\alpha s}
                \int d^3rd^3\rho_1d^3\rho_2 \Psi_\alpha^\ast
                f_{12}^\pi j_0(pr)\Phi_1\Phi_2.
\end{eqnarray}

The matrix element for $\sigma$-meson exchange is
\begin{eqnarray}
  \mathcal{M}_\sigma & = & -i\frac{\Lambda_1}{2}
    \bp\cdot(\beps_1\btimes\beps_2)4\mathcal{W}_\sigma,
    \label{eq:Msigma} \\
  \mathcal{W}_\sigma & = & \frac{\sqrt{2E_\alpha s}}{p}
    \frac{1}{4}\int d^3rd^3\rho_1d^3\rho_2 \Psi_\alpha^\ast
    \frac{1}{2}\sum_{i\neq j}\left(-\nablal_if_{ij}^\sigma
    +f_{ij}^\sigma\nablar_i\right)3j_1(pr)\phat\cdot\rhat
    \Phi_1\Phi_2 \\
  & = & \mathcal{W}_1
    \left(\langle f_{12}^\sigma\rangle+\langle f_{13}^\sigma\rangle+
    \langle{f_{13}^\sigma}'\rangle \right), 
\end{eqnarray}
where $f_{ij}j_0(pr)$ has been replaced by 
$\frac{1}{2p}(-\nablal_if_{ij}+f_{ij}\nablar_i)j_1(pr)$
because of the symmetrization in Eq.~(\ref{eq:Os}). 
The $\langle f_{ij}^x\rangle$ are defined as the averages
\begin{eqnarray}
        \langle f_{12}^x\rangle & = & 
                \frac{1}{\mathcal{W}_1}\sqrt{2E_\alpha s}
                \int d^3rd^3\rho_1d^3\rho_2 \Psi_\alpha^\ast
                f_{12}^xj_0(pr)\Phi_1\Phi_2, \nonumber \\
        \langle f_{13}^x\rangle & = & 
                \frac{1}{\mathcal{W}_1}\frac{\sqrt{2E_\alpha s}}{p}\phat\cdot
                \int d^3rd^3\rho_1d^3\rho_2 \Psi_\alpha^\ast
                f_{13}^x(\nablar_r+\nablar_{\rho_1}-\nablar_{\rho_2})
                3j_1(pr)\phat\cdot\rhat\Phi_1\Phi_2, \nonumber \\
        \langle {f_{13}^x}'\rangle & = & 
                \frac{1}{\mathcal{W}_1}\frac{\sqrt{2E_\alpha s}}{p}\phat\cdot
                \int d^3rd^3\rho_1d^3\rho_2 \Psi_\alpha^\ast
                \left(-\nablal_r-\nablal_{\rho_1}+\nablal_{\rho_2}\right)
                f_{13}^x3j_1(pr)\phat\cdot\rhat\Phi_1\Phi_2,
\label{eq:fijDef}
\end{eqnarray}
where $x$ could be any of the heavy mesons. 
This separates the in-deuteron exchanges $f_{12}$ from exchanges between 
nucleons from different deuterons ($f_{13}$ and $f_{13}'$). 
Exchanges between other pairs of nucleons can be reduced to these two 
because of the symmetries of the $dd$ and $\alpha$ wave functions. 
Some of the necessary integrals are presented in the Appendix.

The $\omega$-exchange form factor is given by
\begin{eqnarray}
        \mathcal{W}_\omega & = & \mathcal{W}_1
       (-\langle f_{12}^\omega\rangle+2\langle f_{13}^\omega\rangle),
\end{eqnarray}
with the same external spin factors as in Eq.~(\ref{eq:Msigma}) and with 
averages defined as in Eq.~(\ref{eq:fijDef}). Due to cancellations, the 
$\langle{f_{13}^\omega}'\rangle$ term cannot contribute.

The $\rho$-exchange amplitude has an isospin factor $\btau_i\cdot\btau_j$ 
that makes it possible to match the initial-state deuterons with the second 
term of the $\alpha$-particle wave function [Eq.~(\ref{eq:alpha})], giving 
sizable exchanges between the deuterons. The large ratio of the $\rho$ tensor 
to vector couplings enhances this amplitude despite the small value of the
$\rho$ coupling constant.
The $\rho$ form factor is
\begin{eqnarray}
        \mathcal{W}_\rho & = & \mathcal{W}_1
                3\left[\langle f_{12}^\rho\rangle+(1+C_\rho)
        (\langle f_{13}^\rho\rangle-\langle{f_{13}^\rho}'\rangle) \right].
\label{eq:Mrho}
\end{eqnarray}

Similarly, the $\rho$-$\omega$--mixing amplitude has a $\tau_i^3\tau_j^3$ term 
that involves the large $\rho$ tensor coupling, which gives a large contribution 
to CSB for $dd\to\alpha\pi^0$ and possibly also for $np\to d\pi^0$.
The form factor for $dd\to\alpha\pi^0$ is
\begin{eqnarray}
        \mathcal{W}_{\rho-\omega} & = & \mathcal{W}_1
                \frac{\Lambda_{\rho-\omega}}{\Lambda_1}
                [(3+C_\rho)\langle f^{\rho\omega}_{13}\rangle
                -(1+C_\rho)\langle{f_{13}^{\rho\omega}}'\rangle],
\label{eq:Mrhomega}
\end{eqnarray}
which follows immediately from the expressions for the $\rho$ and $\omega$ 
exchanges. 
The $f_{12}$ term vanishes since the $1+\tau_1^3\tau_2^3$ term of 
Eq.~(\ref{eq:Orhomega}) gives zero when acting on a deuteron and the 
$\bsigma_1\times\bsigma_2$ term vanishes due to the spin-couplings in the wave 
functions.

The photon exchange contribution is
\begin{eqnarray}
        \mathcal{W}_{\gamma} & = & \mathcal{W}_1
                \frac{\Lambda_{\gamma}}{\Lambda_1}
                [(3+2\lambda_p)\langle f^{\gamma}_{13}\rangle
                -(1+2\lambda_p)\langle{f_{13}^{\gamma}}'\rangle],
\label{eq:Mgamma}
\end{eqnarray}
where again the $f_{12}$ term vanishes. In the simplified model the photon
exchange only occurs between pairs of protons, thus not benefiting from the 
coherence the other amplitudes experience. 
However, the relatively large coupling makes this amplitude important.

Thus all HMECs have contributions to $dd\to\alpha\pi^0$. 
While there are some cancellations of the in-deuteron ($f_{12}$) 
and the derivative ($f_{13}'$) exchange terms between the different heavy 
mesons, all the $f_{13}$ contributions are of the same sign.
Since in all cases $\langle f_{13}\rangle$ is much larger than 
$\langle f_{12}\rangle$ and $\langle f_{13}'\rangle$ 
(see Table~\ref{tab:fij} below), they 
dominate the matrix element and the cross section, adding coherently with 
each other and also with the one-body and pion-exchange terms.
Also the photon graph is of the same sign.
Thus the internal spin-isospin symmetries of the $dd\!:\!\alpha$ system used here  
strongly favor the one-body and meson-exchange amplitudes if the plane
wave approximation is used. 
This will be demonstrated quantitatively in Sec.~\ref{Results}.

\section{Results}
\label{Results}
The matrix elements of the previous section are evaluated numerically using 
the simplified (Gaussian) wave functions. At most a double integration with a 
separate single integral was needed, which was carried out using standard 
Gauss-Legendre techniques.
Explicit formulas for the integrals are presented in the Appendix.
 
The Yukawa averages are tabulated in Table~\ref{tab:fij} for each operator and 
both energies relevant to the IUCF experiment.
\begin{table}[t]
\caption{The Yukawa averages (see text for definitions) evaluated for 
$dd\to\alpha\pi^0$ at the two IUCF energies, using Gaussian wave functions.}
\begin{ruledtabular}
\begin{tabular}{c|ccc|ccc}
Operator & \multicolumn{3}{c|}{$T_d=228.5$~MeV} & 
\multicolumn{3}{c}{$T_d=231.8$~MeV} \\
& $\langle f_{12}\rangle$ & $\langle f_{13}\rangle$ & 
$\langle{f_{13}'}\rangle$ & $\langle f_{12}\rangle$ & $\langle f_{13}\rangle$ 
& $\langle{f_{13}'}\rangle$ \\ \hline
$\pi$           & 0.0172   & 0       & 0        & 0.0172   & 0      & 0 \\
$\sigma$        & 0.0292   & 0.470   & 0.0220   & 0.0292   & 0.490  & 0.0229 \\
$\omega$        & 0.0228   & 0.395   & 0.0106   & 0.0228   & 0.412  & 0.0111 \\
$\rho$          & 0.00095  & 0.0165  & 0.00046  & 0.00095  & 0.0172 & 0.00048\\
$\rho$-$\omega$ & 0.00445  & 0.0704  & 0.00377  & 0.00445  & 0.0734 & 0.00393\\
$\gamma$        & 0.00073  & 0.0032  & 0.00109  & 0.00073  & 0.0033 & 0.00111
\end{tabular}
\end{ruledtabular}
\label{tab:fij}
\end{table}
The $\langle f_{12}\rangle$ contributions are the same at both energies since 
the in-deuteron exchange is independent of the energy in our plane-wave model. 
For completeness the $\langle f_{12}\rangle$ values for $\rho$-$\omega$ and photon
exchanges are given, even though these, as discussed above, do not contribute to 
the matrix element in our simplified model. 
Note that even though the $\rho$-exchange, $\rho$-$\omega$--mixing, and
photon integrals are much smaller than those
for the other meson exchanges, they will be multiplied 
by large constant factors in the definitions of the matrix elements, 
Eqs.~(\ref{eq:Mrho})--(\ref{eq:Mgamma}). 
This drastically increases their relative importance. 

The matrix elements and cross sections calculated from these averages are given
in Table~\ref{tab:matrixelem}, individually for each amplitude and as a grand 
total. 
For comparison purposes, the experimental cross sections are included and the 
matrix elements are given relative to the one-body matrix element.
\begin{table}[t]
\caption{Matrix elements and cross sections evaluated for $dd\to\alpha\pi^0$ 
at the two IUCF energies. 
The matrix elements are given relative to the one-body matrix element, with 
the relevant CSB mechanism indicated. 
The experimental cross sections are also given.}
\begin{ruledtabular}
\begin{tabular}{c|cccc}
Operator (CSB mech.) & $\mathcal{M}(228.5)$ & $\mathcal{M}(231.8)$ & 
$\sigma(228.5)$ & $\sigma(231.8)$  \\
                  & [$\mathcal{M}_1$] & [$\mathcal{M}_1$] & [pb] & [pb] \\ \hline
$\pi$  $(\delta M,\bar\delta M)$   & 0.128  & 0.128  & 0.011        & 0.014 \\
$1$  $(\pi$-$\eta)$                & 1      & 1      & 0.688        & 0.869 \\
$\sigma$  $(\pi$-$\eta)$           & 0.522  & 0.543  & 0.187        & 0.256 \\
$\omega$  $(\pi$-$\eta)$           & 0.766  & 0.801  & 0.404        & 0.557 \\
$\rho$  $(\pi$-$\eta)$             & 0.344  & 0.359  & 0.082        & 0.112 \\
$\rho$-$\omega$  $(\rho$-$\omega)$ & 1.546  & 1.612  & 1.645        & 2.256 \\
$\gamma$  (el.-mag.)               & 1.469  & 1.517  & 1.486        & 1.999 \\
total                              & 5.78   & 5.96   & 23.0         & 30.8 \\
Exp.~\cite{IUCFCSB}                & ---    & ---    & $12.7\pm2.2$ & $15.1\pm3.1$
\end{tabular}
\end{ruledtabular}
\label{tab:matrixelem}
\end{table}
All the heavy meson exchanges are of the same order as the one-body term, with
the $\rho$-$\omega$ mixing being the largest.
Adding all amplitudes gives a total matrix element that is almost six times
that of the one-body, increasing the cross section from a meager 0.69~pb at
$T_d=228.5$~MeV (0.87~pb at 231.8~MeV) to 23~pb (31~pb). 
This is of the same order as the IUCF data $\sigma=12.7\pm2.2$~pb and 
$15.1\pm3.1$~pb.
Note that the cross section is not strictly linear in the pion
momentum --- the momentum transfer in the wave functions introduces a 
dependence on the deuteron momentum, which modifies the linearity, 
at least in this simplified model. 
For example, Eq.~(\ref{sig1}) for the one body term contains 
the square of the deuteron momentum in an exponential.

The relative proportions of the pion-exchange 
($\delta M-\frac12\bar\delta M$), photon exchange, \mbox{$\rho$-$\omega$--},
and $\pi$-$\eta$--mixing (sum of one-body and HMEC) contributions 
to the matrix element are roughly 
$\pi$:$\gamma$:$\rho$-$\omega$:$\pi$-$\eta$=1:11:12:21. 
Thus the formally leading seagull terms make up only about 2\% of the total 
matrix element.
The total cross section can be expressed in terms of the relative 
contributions of the different CSB mechanisms such that
the dependences on the corresponding parameters are made explicit:
\begin{eqnarray}
  &&\sigma(228.5~{\rm MeV}) = (23.0~{\rm pb})
  \left(0.254
    +0.0188\frac{\delta M}{2.03~{\rm MeV}}
    +0.0034\frac{\bar\delta M}{-0.74~{\rm MeV}} \right. \nonumber \\
    \lefteqn{ + \left. 
    0.456\frac{g_{\eta NN}}{\sqrt{4\pi\cdot0.51}}
      \frac{\langle\eta|H|\pi^0\rangle}{(-4200~{\rm MeV^2})}
    +0.268\frac{g_\rho g_\omega}{4\pi\sqrt{0.43\cdot10.6}}
      \frac{\langle\omega|H|\rho^0\rangle}
           {(-4300~{\rm MeV^2})}\right)^2.} 
\label{eq:fracs}
\end{eqnarray}
Here the numerical coefficients are the fractions of the 
matrix element belonging to each of the considered mechanisms, 
assuming our choice of parameter values. 
The various terms are normalized, as indicated, to these values.
The photon exchange diagram is represented by just a number, 
since its parameters are well-known.
Note that the second and third terms are constrained by the neutron-proton 
mass difference [Eq.~(\ref{eq:M})].
The $\eta$-$\pi^0$--mixing term can be further separated to show the relative 
contribution of the various HMEs. Thus,
\begin{eqnarray}
  \sqrt{\sigma_{\pi\eta}} & = & \left(2.18\ \sqrt{\rm pb}\right) 
  \left(0.380+0.198\frac{g_\sigma^2}{4\pi\cdot7.1}+
  0.291\frac{g_\omega^2}{4\pi\cdot10.6}+
  0.131\frac{g_\rho^2}{4\pi\cdot0.43}\right), 
\label{eq:pietafracs}
\end{eqnarray}
where $\sigma_{\pi\eta}$ is the cross section from $\pi$-$\eta$ contributions 
alone and the first number in the second parenthesis is the one-body contribution.
At the higher IUCF energy the relative weights of the different 
contributions in Eqs.~(\ref{eq:fracs}) and (\ref{eq:pietafracs}) 
remain more or less the same, with only minor changes.
The sensitivity of the cross section calculation to a different choice 
of couplings can easily be found from these two formulas. 
For example, using instead the large $g_\eta^2/4\pi=3.68$ and 
$\langle\eta|H|\pi^0\rangle=-5900$~MeV$^2$ as in Ref.~\cite{vKNM}, the 
cross section increases from 23 to 118~pb (31 to 158~pb at 231.8~MeV).

\section{Discussion}
\label{Disc}
Our simplified model  keeps a complete treatment of the dominant pieces of the 
spin-isospin couplings in the bound state wave functions, even though it ignores 
some dynamics of the $dd\to\alpha\pi^0$ reaction and the distortion of the 
initial state. 
As a result, the  symmetries of the bound state wave functions allow us to 
determine the  CSB amplitudes that are guaranteed to be important for a 
full calculation. 

This treatment shows that the LO $\pi$-rescattering term 
(from the the seagull interactions) is suppressed because of a poor overlap 
with the initial- and final-state wave functions. 
Photon loops, at NLO, vanish due to symmetries 
and cancellations, while a three-body contribution might survive, but has not 
yet been calculated. 
On the other hand, the NNLO one-body amplitude and N$^4$LO heavy-meson 
exchanges are strongly favored, adding coherently with each other. 
Their dominance would be even more spectacular if a larger value for the 
$\eta NN$ coupling were used.
Also the $\rho$-$\omega$--mixing and photon-exchange terms are important and 
enter at the same level as the one-body and HMEC terms.

We note that our analysis assumes that pions are produced in $s$-waves,
as one would expect for a near-threshold reaction. 
This is supported by the IUCF experiment, where the energy dependence is 
consistent with $s$-wave production~\cite{IUCFCSB}.

If our simple wave functions and the plane wave approximation are used, 
then, within the resonance saturation picture, the dominant CSB mechanisms for the 
$dd\to\alpha\pi^0$ reaction are identified with $\pi$-$\eta$ mixing 
(one-body enhanced by HMECs), followed by $\rho$-$\omega$ mixing and photon 
exchange, and finally a small contribution from pion rescattering (related to 
the neutron-proton mass difference).

The coherent sum of these contributions leads to a cross section of
the same order of magnitude as the observed one~\cite{IUCFCSB}, but more 
needs to be done before making a detailed assessment of the quality of
the agreement between theory and experiment. 

It is likely that the relative importance of these amplitudes will be shifted 
once realistic wave functions are used. 
Preliminary calculations suggest that the one-body term can in fact be enhanced 
by as much as a factor of three or four with a realistic $\alpha$-particle 
wave function. 
This is expected, since this amplitude is sensitive to the high-momentum tail 
of the wave function, which is very small for a Gaussian.
Preliminary estimates also show  that spin-dependent initial state 
interactions  enhance the pion rescattering contribution. 
We stress that the HMECs are less sensitive to the $\alpha$-particle wave function
and should remain crucial for the interpretation of CSB in the
$dd\to\alpha\pi^0$ reaction. 

A full model calculation, using realistic bound-state and $dd$-scattering 
wave functions is needed in order to have a clear understanding of the 
CSB mechanisms behind the $dd\to\alpha\pi^0$ reaction. 
Furthermore, the effects of CSB in the wave functions and some diagrams 
ignored here, such as the long-range part of the various NNLO pion loops and the 
N$^3$LO recoil part of the $\Delta$-excitation term, should  be included. 
In particular, it is necessary to include the non--z-graph part of photon exchange,
e.g., the Coulomb interaction in the initial and final states 
(Fig.~\ref{fig:wfcsb}).
Such an investigation is currently in progress and will be reported later.
The general conclusions and insights from the present paper provide 
important guidelines for that work.

We note a very interesting parallel between the $dd\to\alpha\pi^0$ process 
considered here, and the reaction $pp\to pp\pi^0$.
In both cases, a formally-leading diagram is suppressed and 
the sub-leading diagrams are crucial to explain the cross section.
Despite several serious efforts that have yielded substantial insights 
into the various $NN\to NN\pi$ systems, the $pp\to pp\pi^0$ reaction is still 
not completely understood, especially regarding spin observables~\cite{pppi0}. 

Higher-order interactions --- such as the heavy-meson-exchange terms, which 
could increase the role of $\pi$-$\eta$ mixing --- might help improve the 
agreement between the TRIUMF result for $A_{\rm fb}(np\to d\pi^0)$~\cite{Allena} 
and theoretical estimates based on reasonable values for $\delta M$ and 
$\bar\delta M$~\cite{vKNM}.
It is thus necessary that a future calculation of $A_{\rm fb}(np\to d\pi^0)$ 
includes these higher-order terms.
The three reactions $pp\to pp\pi^0$, $np\to d\pi^0$, and $dd\to\alpha\pi^0$
provide important testing grounds for any pion-production 
model that intends to include effects beyond-leading-order.

\begin{acknowledgments}
We are grateful to Andrew Bacher, Edward Stephenson and Allena Opper 
for encouragement, many useful discussions, and providing results from
the IUCF and TRIUMF experiments prior to publication. 
We thank Ulf-G. Mei{\ss}ner for useful discussions.
We are grateful to the National Institute for Nuclear Theory at the 
University of Washington for its hospitality and for arranging CSB workshops
where part of this work was completed.
U.v.K. is grateful to the Nuclear Theory Group at the University of Washington
for its hospitality, and to RIKEN, Brookhaven National Laboratory and 
the U.S. Department of Energy [DE-AC02-98CH10886] for providing the facilities
essential for the completion of this work. 
G.A.M. thanks the ECT* (Trento), the INT (UW, Seattle), and the CSSM 
(Adelaide) for providing hospitality during the completion of this work.
This work was supported in part by the
Magnus Ehrnrooth Foundation (J.A.N.), 
the grant POCTI/FNU/37280/2001 (A.C.F),
the NSF grant NSF-PHY-00-70368 (A.G.),
the DOE grants DE-FG02-87ER40365 (C.J.H.), 
DE-FG02-93ER40756 and DE-FG02-02ER41218 (A.G.), 
DE-FC02-01ER41187 and DE-FG03-00ER41132 (A.N.),
DE-FG-02-97ER41014 (G.A.M.),
and DE-FG03-01ER41196 (U.v.K.), 
and by an Alfred P. Sloan Fellowship (U.v.K.).  
\end{acknowledgments}

\appendix*
\section{Explicit expressions for Yukawa averages}
\label{app:yukawa}
The averages of the different Yukawa factors of Eq.~(\ref{eq:fijDef}) can be 
reduced to at most two-dimensional integrals, using the Gaussian wave 
functions of Sec.~\ref{sec:Model}. 
The angular and one of the radial integrals can be carried out
analytically, resulting in the explicit formulas
\begin{eqnarray}
  \langle f_{12}^x\rangle & = & 
    \frac{1}{\mathcal{F}_1}
    \int d\rho_1\rho_1^2f_{12}^x\expup{-\frac{\rho_1^2}{\gamma^2}} 
    \int d\rho_2\rho_2^2\expup{-\frac{\rho_2^2}{\gamma^2}} 
    p\int drr^2j_0(pr)\expup{-\frac{2r^2}{\alpha^2}}, \nonumber \\
  \langle f_{13}^x\rangle & = & 
    \frac{1}{\mathcal{F}_1}\int d\rho\rho^2\expup{-\frac{2\rho^2}{\gamma^2}}
    \left\{p\gamma^2
    \int dr rj_0(pr)\expup{-\frac{2r^2}{\alpha^2}}
    \int dr_{13}r_{13}f_{13}^x
    \left(\expup{-\frac{2(r_{13}-r)^2}{\gamma^2}}-
    \expup{-\frac{2(r_{13}+r)^2}{\gamma^2}}\right)\right. \nonumber \\
    & - & \frac{2\gamma^2}{\beta^2}
    \int dr j_1(pr)\expup{-\frac{2r^2}{\alpha^2}}
    dr_{13}r_{13}f_{13}^x \nonumber \\
   & \times & \left.\left[\left(rr_{13}-\frac{\gamma^2}{4}-r^2\right)
     \expup{-\frac{2(r_{13}-r)^2}{\gamma^2}}+
     \left(rr_{13}+\frac{\gamma^2}{4}+r^2\right)
     \expup{-\frac{2(r_{13}+r)^2}{\gamma^2}}\right]\right\}, \nonumber \\
  \langle {f'}_{13}^x\rangle & = & 
    \frac{1}{\mathcal{F}_1}\frac{4\gamma^2}{\alpha^2}
    \int d\rho\rho^2\expup{-\frac{2\rho^2}{\gamma^2}}
      \int dr j_1(pr)\expup{-\frac{2r^2}{\alpha^2}}
      dr_{13}r_{13}f_{13}^x \nonumber \\
    & \times & \left[\left(rr_{13}-\frac{\gamma^2}{4}\right)
      \expup{-\frac{2(r_{13}-r)^2}{\gamma^2}}+
      \left(rr_{13}+\frac{\gamma^2}{4}\right)
      \expup{-\frac{2(r_{13}+r)^2}{\gamma^2}}
      \right] , \nonumber \\
  \mathcal{F}_1 & = & \left[\int d\rho\rho^2
    \expup{-\frac{\rho^2}{\gamma^2}}\right]^2 
    p\int dr r^2j_0(pr)\expup{-\frac{2r^2}{\alpha^2}},
\end{eqnarray}
where $r_{13}=|\ri{1}-\ri{3}|$ and $1/\gamma^2=1/\alpha^2+1/(2\beta^2)$.

\bibliographystyle{unsrt}

\end{document}